\documentclass[fleqn,useAMS,usenatbib]{mnras}
\usepackage{amssymb}
\usepackage{amsmath}
\usepackage{xspace}
\usepackage{longtable}
\usepackage{graphicx}
\usepackage{times}
\newcommand{\HI}{\mbox{\sc H{i}}}

\newcommand{\kms}{\mbox{km~s$^{-1}$}}
\newcommand{\msun}{\mbox{$\rm{M}_\odot$}}

\title[WALLABY]
  {WALLABY pre-pilot survey: Two dark clouds in the vicinity of NGC~1395}
\author[Wong, O.I. et al.]{O.I.~Wong,$^{1,2,3}$  A.R.H.~Stevens,$^{2,3}$ B.-Q.~For,$^{2,3}$ T.~Westmeier,$^{2,3}$   M.~Dixon,$^{4}$ S.-H. Oh,$^{5}$ G.I.G.~J\'ozsa,$^{6,7,8}$
\newauthor T.N.~Reynolds,$^{2,3}$  K.~Lee-Waddell,$^{2,1}$ J.~Rom\'{a}n,$^{9,10,11}$  L.~Verdes-Montenegro,$^{9}$ H.M.~Courtois,$^{12}$,  
\newauthor  D.~Pomar\`{e}de,$^{13}$ C.~Murugeshan,$^{1,4,3}$ M.T. Whiting,$^{14}$ K.~Bekki,$^2$ F.~Bigiel,$^{8}$  A.~Bosma,$^{15}$ B.~Catinella,$^{2,3}$  
\newauthor  H.~D\'{e}nes,$^{16}$  A.~Elagali,$^{2,17}$   B.W.~Holwerda,$^{18}$ P.~Kamphuis,$^{19}$  V.A.~Kilborn,$^{4,3}$  D.~Kleiner,$^{20}$  
\newauthor  B.S.~Koribalski,$^{14,21}$ F.~Lelli,$^{22}$  J.P.~Madrid,$^{23}$  K.B.W.~McQuinn,$^{24}$   A.~Popping,$^{2,25}$ J.~Rhee,$^{2,3}$ 
\newauthor S.~Roychowdhury,$^{2,3}$  T.C.~Scott,$^{26}$ C. Sengupta,$^{27}$ K.~Spekkens,$^{28}$ L.~Staveley-Smith,$^{2,3}$ \&  B.P.~Wakker$^{29}$ \\ 
$^1$ATNF, CSIRO, Space \& Astronomy, PO Box 1130, Bentley, WA 6102, Australia\\
$^2$International Centre for Radio Astronomy Research, University of Western Australia, 35 Stirling Highway, Crawley, WA 6009, Australia\\
$^3$ARC Centre of Excellence for All-Sky Astrophysics in 3 Dimensions (ASTRO 3D), Australia\\
$^4$Centre for Astrophysics and Supercomputing, Swinburne University of Technology, Hawthorn, Victoria 3122, Australia\\
$^5$Department of Physics and Astronomy, Sejong University, 209 Neungdong-ro, Gwangjin-gu, Seoul, Republic of Korea\\
$^6$South African Radio Astronomy Observatory, 2 Fir Street, Black River Park, Observatory, Cape Town, 7925, South Africa\\
$^7$Department of Physics and Electronics, Rhodes University, PO Box 94, Makhanda, 6140, South Africa\\
$^8$Argelander-Institut f\"ur Astronomie, Universit\"at Bonn, Auf dem H\"ugel 71, 53121 Bonn, Germany\\
$^{9}$Instituto de Astrofísica de Andalucía, CSIC, Glorieta de la Astronomía, E-18080, Granada, Spain\\
$^{10}$Instituto de Astrof\'{\i}sica de Canarias, c/ V\'{\i}a L\'actea s/n, E-38205, La Laguna, Tenerife, Spain\\
$^{11}$Departamento de Astrof\'{\i}sica, Universidad de La Laguna, E-38206, La Laguna, Tenerife, Spain\\
$^{12}$Univ Lyon, Univ Claude Bernard Lyon 1, IUF, IP2I Lyon, F-69622, Villeurbanne, France\\
$^{13}$Institut de Recherche sur les Lois Fondamentales de l'Univers, CEA Universit\'{e} Paris-Saclay, France\\
$^{14}$CSIRO, Space \& Astronomy, PO Box 76, Epping, NSW 1710, Australia\\
$^{15}$Aix Marseille Univ, CNRS, CNES, LAM, Marseille, France\\
$^{16}$ASTRON, The Netherlands Institute for Radio Astronomy, Dwingeloo, the Netherlands\\
$^{17}$Telethon Kids Institute, Perth Children’s Hospital, Perth, Australia\\
$^{18}$University of Louisville, Department of Physics and Astronomy, 102 Natural Science Building, 40292 KY Louisville, USA\\
$^{19}$Ruhr University Bochum, Faculty of Physics and Astronomy, Astronomical Institute, 44780 Bochum, Germany\\
$^{20}$INAF – Osservatorio Astronomico di Cagliari, Via della Scienza 5, 09047 Selargius (CA), Italy\\
$^{21}$School of Science, Western Sydney University, Locked Bag 1797, Penrith, NSW 2751, Australia\\
$^{22}$INAF - Arcetri Astrophysical Observatory, Largo Enrico Fermi 5, I-50125, Florence, Italy\\
$^{23}$Departmento de F\'isica y Astronom\'ia, La Universidad de Tejas de el Valle del R\'io Grande, Brownsville, TX 78520, USA\\
$^{24}$Rutgers University, Department of Physics and Astronomy, 136 Frelinghuysen Road, Piscataway, NJ 08854, USA\\
$^{25}$TMC Data Science, High Tech Campus 96, 5656 AG Eindhoven, the Netherlands\\
$^{26}$Institute of Astrophysics and Space Sciences (IA), Rua das Estrelas, 4150--762 Porto, Portugal\\
$^{27}$Purple Mountain Observatory (CAS), No. 8 Yuanhua Road, Qixia District, Nanjing 210034, China\\
$^{28}$Department of Physics and Space Science, Royal Military College of Canada,  PO Box 17000, Station Forces, Kingston, Ontario, Canada K7K 7B4\\
$^{29}$University of Wisconsin-Madison, 475 N Charter St, Madison, WI 53706, USA\\
}
\date{Released 2021 December 01}
\pagerange{\pageref{firstpage}--\pageref{lastpage}} \pubyear{2021}

\def\LaTeX{L\kern-.36em\raise.3ex\hbox{a}\kern-.15em
    T\kern-.1667em\lower.7ex\hbox{E}\kern-.125emX}

\begin{document}

\label{firstpage}

\maketitle
\begin{abstract}  

We present the Australian Square Kilometre Array Pathfinder (ASKAP) WALLABY 
pre-pilot observations of two `dark' \HI\ sources (with \HI\ masses of a few times 10$^8$~\msun\ and no known stellar counterpart)
that reside within 363~kpc of NGC~1395, the most massive early-type galaxy in  
the Eridanus group of galaxies.  We investigate whether these `dark' \HI\ sources
have resulted from past tidal interactions or whether they are an extreme class of low surface
brightness galaxies.  Our results suggest that both scenarios are possible, and not mutually exclusive.
The two `dark'  \HI\ sources are compact, reside
in relative isolation and are more than 159~kpc away from their nearest \HI-rich galaxy neighbour.
Regardless of origin, the \HI\ sizes and masses of both `dark' \HI\ sources are consistent with the \HI\ size--mass
relationship that is found in nearby low-mass galaxies, supporting the possibility that these
\HI\ sources are an extreme class of low surface brightness galaxies.  We identified three analogues of
candidate primordial `dark' \HI\ galaxies within the TNG100 cosmological, hydrodynamic simulation.  All
three model analogues are  dark matter-dominated,  have assembled most of their mass 12--13 Gyr ago, and have not 
experienced much evolution until cluster infall  1--2 Gyr ago.  Our WALLABY pre-pilot science results
suggest that the upcoming large area \HI\ surveys will have a significant impact on our
understanding of low surface brightness galaxies and the physical processes that shape them.

\end{abstract}


\section{Introduction} 

%While the star formation histories of galaxies are well-understood 
%through studies such as those of the star-forming `main sequence' and
% the cosmic star formation rate densities, the  formation of
% early-type galaxies (ETGs or spheroidals) from spirals is still an 
%active area of 
%research \citep[e.g.\ ][]{spitzer51,sandage61,vandenbergh76,kormendy12,martin18,serra19,tacchella19,trayford19,jackson19,park19}.  There are still several open questions in the field of
%galaxy transformation, with respect to the build-up of spheroidal
%galaxies. 

%Observations from integral field spectroscopic surveys have 
%been largely responsible for providing our current understanding that
%the vast majority of early-type or spheroidal galaxies are rotation velocity
%-dominated, while the true number of dispersion velocity-dominated 
%elliptical galaxies are much fewer \citep[e.g.\ ][]{emsellem11,cappellari11}.

Previous \HI\ large area surveys have revealed `dark galaxies' or \HI-dominated systems with no optical counterparts to be very rare \citep{kilborn00,ryder01,kilborn06,haynes07,wong09,giovanelli10,adams13,oosterloo13,serra15,madrid18,lee19}.  These \HI-dominated systems are often remnants of past tidal (or hydrodynamic) interactions and/or could be more extreme cousins of very low surface brightness galaxies such as the class of \HI-rich tidal dwarf galaxies \citep{bournaud07,lelli15,lee16,leisman16,sengupta17} and `almost-dark' galaxies \citep[e.g.\ ][]{leisman17,sengupta19,janowiecki19,scott21}.  Galaxy evolution models have demonstrated that the majority of \HI\ clouds without optical counterparts are likely the result of galaxy-galaxy interactions \citep[e.g.\ ][]{bekki05,duc08,taylor17} and have evolved from tidal streams. Currently, large extended \HI\ streams are also relatively rare and are typically discovered in galaxy groups and clusters in the nearby Universe \citep[e.g.\ ][]{schneider83,oosterloo05,haynes07,serra13,lee-waddell14}.  Recently, the discovery of \HI\ tidal tails near NGC~1316 (also known as Fornax A) further constrained  the favoured merger formation scenario for NGC~1316 that was previously proposed \citep{iodice17,serra19}.  This suggests that our understanding of the typical early-type galaxy (ETG) formation scenario may improve if \HI\ tails are found to be more common.

%\citet{taylor20} estimates a stripped cloud dissolution rate of 1--10 solar masses per year within the Virgo cluster environment.
%At present, large extended \HI\ streams are relatively rare and are typically discovered in galaxy groups and clusters in the nearby Universe \citep[e.g.\ ][]{oosterloo05,haynes07,serra13,lee-waddell14}.  Recently, the discovery of \HI\ tidal tails near NGC~1316 (also known as Fornax A) further constrained and changed the favoured merger formation scenario for NGC~1316 that was previously proposed \citep{iodice17,serra19}.  This suggests that our understanding of the typical early-type galaxy (ETG) formation scenario may change if \HI\ tails are found to be more common.

While star formation is relatively well-understood in massive disk galaxies, the process of star formation in low surface brightness galaxies (and at its most extreme, `almost-dark' or \HI-rich ultra-diffuse galaxies) is still an area of active research \citep{roychowdhury09,hunter11,roychowdhury11,elmegreen15,roychowdhury15}.  Using the Australian Square Kilometre Pathfinder \citep[ASKAP; ][]{johnston07,hotan21}, next-generation \HI\ all-sky surveys such as the Widefield ASKAP L-band Legacy All-sky Blind surveY \citep[WALLABY; ][]{koribalski20} will enable astronomers to further understand the formation and evolutionary processes at play in this extreme class of sources.

%Our current understanding suggests that ETGs can evolve from disk galaxies through both secular and environmental processes \citep[e.g.\ ][]{emsellem11,schawinski14,kaviraj13,jackson19,trayford19}.   Approximately forty to fifty percent of recent spheroidal mass appears to have been assembled via mergers \citep{kaviraj13,martin18,tacchella19,trayford19} whereby 10:1 mergers are thought to be a probable evolutionary pathway for disk-to-spheroid transformations \citep{trayford19} through the growth of the kinematic spheroid, as opposed to the destruction of the disk \citep{clauwens18}. Such merger scenarios are consistent with the accretion of ex-situ stars at large radii to the parent galaxy \citep{genel18} and the finding that gas-poor mergers typically result in loss of angular momenta \citep{lagos17}.  

Taking advantage of the large ASKAP field-of-view, WALLABY early-science observations investigated galaxy interactions in galaxy groups as well as the impact of the galaxy group environment on the gas content and star formation of individual group galaxies \citep{serra15,reynolds19,lee19,elagali19,kleiner19,for19}.  Prior to the start  of the ASKAP pilot survey period, the ASKAP WALLABY team obtained test observations of the Eridanus (NGC~1395) group (or subcluster) of galaxies.  NGC~1395, the most massive ETG in the Eridanus group, has shell-like stellar structures, indicative of mergers and accretion events in the past several Gyrs \citep{malin80,quinn84}. Stellar population analyses of the NGC~1395 globular clusters provide evidence for the active accretion of low-mass satellites \citep{escudero18}. 

The Eridanus group (or subcluster) is part of the Eridanus `supergroup', which is composed of  three groups  of galaxies that will eventually merge into a single cluster with a total mass of approximately $7 \times 10^{13}$~M$_{\odot}$ \citep{brough06}.  The Northern NGC~1407 subcluster is the most evolved subcluster with a significant X-ray halo --- into which the Eridanus group (to the South of NGC~1407) is currently infalling  \citep{willmer89,osmond04,omar05a,brough06}.  The third subcluster, the NGC~1332 galaxy group (South-West of NGC~1407 and West of the Eridanus group) is the least evolved group with the largest fraction of spirals and irregular galaxies. %The NGC~1332 subcluster is at an earlier stage of infall towards the NGC~1407 subcluster, relative to that between the Eridanus subcluster and the NGC~1407 subcluster \citep{brough06}.
To distinguish between the Eridanus supergroup/cluster and the NGC~1395 group (which is also known as the Eridanus group in previous publications), we refer to the NGC~1395 subcluster as the Eridanus group in this paper.

The Eridanus group was targeted by the WALLABY team in order to investigate the processes that occur in galaxies that reside in an in-falling subcluster. Such processes contribute towards the `pre-processing' of galaxies, with  the quenching of star formation in approximately a third of cluster galaxies estimated to be attributable to  `pre-processing' \citep[e.g.\ ][]{jung18}. Further details of these ASKAP pre-pilot test observations can be found in \citet{for20}. A summary of the properties of the Eridanus group can be found in Table~\ref{genprop}. Throughout this paper we adopt the distance of 21~Mpc for the Eridanus group following \citet{for20}.  

In this paper, we investigate the properties of two optically-dark \HI\ sources located within 363~kpc in projection of NGC~1395. These `dark' \HI\ sources, BW~033910-2223 (WALLABY J033911-222322, hereafter known as C1) and BW~033732-2359 (WALLABY J033723-235753, hereafter known as C2), have \HI\ velocities within ~$250$~\kms\ of the central velocity (1638~\kms; Table~\ref{genprop}) of the NGC~1395 group.  Both C1 and C2 were previously detected in the Basketweave survey (BW) using the 64-m Parkes radio telescope \citep{waugh05}.  However, only C2 was correctly identified to be an \HI\ source without an optical counterpart. C1 was previously cross-matched with the optical galaxy, NGC~1403 \citep{doyle05,waugh05}.  Additional follow-up synthesis observations were made by \citet{waugh05} using the Australia Telescope Compact Array (ATCA) in order to obtain more accurate positions of the \HI\ sources found from the Basketweave single-dish observations.

The identification of optical counterparts to C1 and C2 is hindered by the projected chance alignment between both C1 with a background galaxy (NGC~1403), and C2 with a foreground Galactic star.  To determine the connection between C1 and NGC~1403, we obtained new optical spectroscopy of NGC~1403 to verify its recessional velocity that was measured more than a decade ago \citep[e.g.\ ][]{davies87}. While large galaxy clusters are known to have velocity dispersions that are greater than 1000~\kms, this is not expected for less massive galaxy groups such as the Eridanus group.

\begin{table}
\caption{Properties of the Eridanus group.}
\label{genprop}
\footnotesize{
\begin{center}
\begin{tabular}{lcc}
\hline
\hline
Property &   Measurement       & Reference$^*$ \\
\hline
Right Ascension (J2000) & 03:38:32.0   & 1\\
Declination (J2000) & -22:18:51.0   & 1\\
Distance & 21.0~Mpc & 1, 2 \\
Velocity & 1657$\pm 4$~km~s$^{-1}$ & 1 \\
Virial mass & 2.1$\pm 2.7 \times 10^{13}$~M$_{\odot}$ &1 \\
Maximal radial extent & 0.8~Mpc & 1\\
$L_{\rm{Xray}}(0.1-2.0~{\rm{keV}})$ & $6.8 \times 10^{40}$~ergs s$^{-1}$ & 3\\
\hline                     
\end{tabular}    
\end{center}
$^*$Reference key: 1 -- \citet{brough06}; 2--\citet{for20}; 3--\citet{omar05a}  }              
\end{table}

We describe our new observations and data processing in Section~\ref{sectiondata}. Section~\ref{sectionresults} presents  the optically-dark \HI\ clouds that we find within 363~kpc of NGC~1395.  We present the ASKAP \HI\ properties of C1 and C2 in Section~\ref{hiclouds}.  Using deep optical imaging from the Dark Energy Camera Legacy Survey Data Release 9 \citep{dey19}, we investigate the presence of optical (stellar) counterparts to C1 and C2 in Section~\ref{dark}. Section~\ref{sectiondisc} investigates the origin and formation history of these \HI\ clouds.    A summary of our results is presented in Section~\ref{sectionsum}.  For consistency with the parent catalogue \citep{for20}, we adopt a $\Lambda$ Cold Dark Matter ($\Lambda CDM$) cosmology where $\Omega_{\rm{M}} =0.27$, $\Omega_{\rm{\Lambda}} =0.73$ and $H_{\rm{0}}=73$~km~s$^{-1}$~Mpc$^{-1}$.  We note that the results of this paper are not sensitive to the assumed cosmology as the Eridanus group is part of the Local Universe.

%%%%%%%%%%%%%%%%%%%%%%%%%%%%%%%%%
\section{Data}
\label{sectiondata}

\subsection{ASKAP observations}
\begin{figure}
\includegraphics[scale=.23,angle=90]{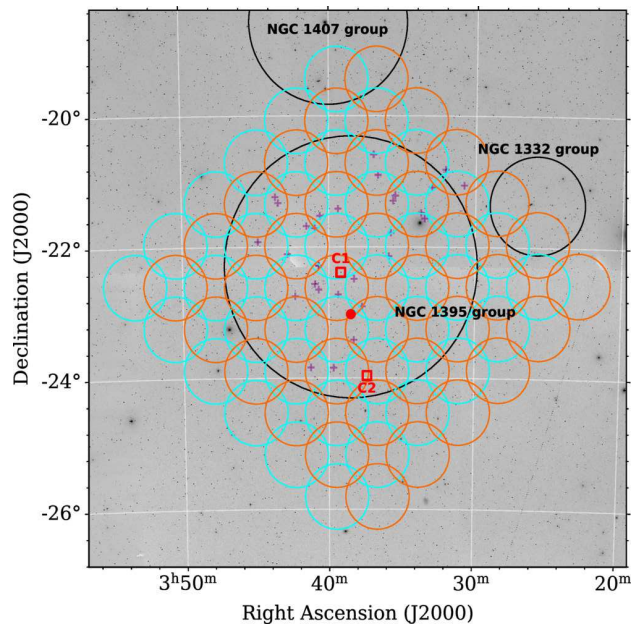}\\
\caption{The diamond-shaped region mapped by Footprint A (cyan) and Fooprint B (orange) of the Eridanus Cluster using ASKAP, overlaid on an optical map from the Digitized Sky Survey.  
Each footprint consists of 36 digitally-formed beams.   The black circles mark the maximum radial extents of the three galaxy groups that make up the Eridanus Cluster \citep{brough06}.  The red dot marks the position of NGC~1395, the largest elliptical galaxy of the Eridanus group.  The red squares mark the location of the two \HI\ clouds (C1 and C2)  with no obvious optical counterparts that are the subject of this paper. The  purple crosses mark the location of galaxies that belong to the Eridanus group \citep{brough06}.  The coordinates and recessional velocity of these galaxies can be found in Appendix~\ref{app-members}. }
\label{footprint}
\end{figure}

Towards the end of the official ASKAP Early Science period (when 31 antennae are correlated together across a bandwidth of 288~MHz) and before the commencement of the official ASKAP pilot surveys, we obtained pre-pilot test observations of the Eridanus group region (see Figure~\ref{footprint}). These test observations were conducted on behalf of the WALLABY survey \citep{koribalski20}.  As the Eridanus group is extended in the North-South direction, we optimise the coverage of this region through two interleaving diamond-shaped footprints whereby the standard $6 \times 6$ square footprint of 36 beams \citep{reynolds19,lee19,elagali19,kleiner19,for19} is rotated by $45\deg$.  The standard calibrator, PKS~1934-638, is used to calibrate the gains and bandpass in these observations.  See Table~\ref{obstab} for more details of these observations. 

\begin{table}
\caption{Summary of the ASKAP observations on the 13 March 2019.}
\label{obstab}
\begin{footnotesize}
\begin{center}
\begin{tabular}{ccccc}
\hline
\hline
Footprint & Centre  & SBID & Integration  & Bandwidth    \\
     & (J2000) &           & (hours) & (MHz) \\
(1)  & (2)       & (3)     & (4)  & (5)           \\
\hline
 A  & 03:39:30.0, $-$22:38:00 & 8168 & 5.8 & 288  \\
 B  & 03:36:44.5, $-$22:37:55 & 8170 & 5.0  & 288  \\
\hline
\end{tabular}
\end{center}
\raggedright Column (1): Footprint ID; column (2): central right ascension and declination of footprint (J2000); column (3): ASKAP scheduling block ID; column (4): integration time in hours; column (5): instantaneous bandwidth of observations. % column (7): The range of RMS values from the 36 beams in each ASKAP pointing.  It should be noted that the RMS  increases as a function of frequency, and varies from beam to beam (see \citet{for20} for more details).              
\end{footnotesize}
\end{table}

\subsection{ASKAP data processing and verification}

Following \citet{for20} and \citet{murugeshan20}, the data in this paper is based on the ASKAP correlations with baselines
shorter than 2~km. Due to the significant levels of radio frequency interference (RFI) in the lower 144~MHz of the 288~MHz bandpass,
only the high frequency half of the band of observations was calibrated and imaged using the  {\sc{AskapSoft}} pipeline
\citep{whiting17}\footnote{\url{http://www.atnf.csiro.au/computing/software/askapsoft/sdp/docs/current/index.html}}.  
The image cubes are produced using the Robust $+0.5$ weighting and are tapered so as to obtain a synthesised beam of $30\arcsec \times 30\arcsec$. The phase and leakage corrections are made using the sky model, phase self-calibration method. The continuum subtraction is conducted in two steps: first in the visibility plane, then in the final image plane before applying the necessary primary beam corrections. The final image cubes for each footprint are then mosaicked together in the image plane, beam by beam, to form the final 72-beam two-footprint observations of the Eridanus group.  The default 18.5~kHz spectral resolution of the final image cubes results in channel widths that are $\sim$4~km~s$^{-1}$ at the frequency relevant to \HI\ in the Eridanus group.  The calibrated imaging products resulting from these observations can be downloaded from the CSIRO ASKAP Science Data Archive (CASDA).

The \HI\ sources were automatically identified using the SoFiA  source finder \citep[version 2.0.1; ][]{sofia,westmeier20}. We refer the reader to \citet{for20} for the complete description of the \HI\ source catalogue  mapped by these ASKAP pre-pilot observations. We mark the location of the two `dark' \HI\ sources that are the focus of this paper as open squares in Figure~\ref{footprint}.  The \HI\ source north of NGC~1395 is  WALLABY~J033911-222322 (hereafter known as C1) and the \HI\ source south of NGC~1395 is  WALLABY~J033723-235745 (hereafter known as C2).

Previous observations of the same region from the Parkes Basketweave survey (BW) and its Australia Telescope Compact Array (ATCA) follow-up survey \citep{waugh05} enable a detailed verification study of these ASKAP pre-pilot observations. The Parkes BW observations reached an RMS of 7 mJy~beam$^{-1}$, and the follow-up ATCA observations had a median RMS of 6.5~mJy  with a synthesised beam of approximately 113\arcsec\ by 296\arcsec. We refer the reader to \citet{for20} for the complete description of this verification study and the data quality assessment of these pre-pilot observations.  We find that the RMS values of these ASKAP pre-pilot observations vary across the field and range between 2.4 to 4.4~mJy. The central beams typically have lower RMS than the beams at the edge of the field.

\newpage \noindent     Following  \citet{for20}, we include the additional 20~percent flux correction factor to the measured \HI\ integrated fluxes ($S_{\rm{HI}}$) in this paper.

\subsection{Optical spectroscopy}
In addition to the ASKAP observations, we  obtained low resolution optical spectroscopy of NGC~1403 using the WiFeS instrument on the 2.3~m telescope at Siding Springs Observatory on the 30 November 2020 through Directors Discretionary Time allocation. The purpose of these observations was to verify the recessional velocity of NGC~1403 that had been measured more than a decade ago \citep[e.g.\ ][]{davies87,paturel02}. 

Using the B3000 grating and the RT560 beam splitter, we obtained a total on-source exposure time of 120~minutes, which involved three pairs of 40~minute on-source exposures, followed by a 15-minute offset sky exposure. Standard calibration exposures were also obtained in addition to a 15-minute exposure of the standard star, GD71.   The spectral resolution from the B3000 grating is approximately 2.2~\AA\ and spans the wavelength range of 3600~\AA\ to 5700~\AA\ \citep{dopita10}. These observations were calibrated and processed using the {\sc{PyWifes}} reduction pipeline \citep{childress14}.  The recessional velocity of the calibrated spectrum was estimated using {\sc{Marz}} by fitting to an early-type galaxy template \citep{hinton16}.

\begin{table}  
\caption{Comparing the ASKAP observations of the two \HI\ clouds to previous Basketweave single-dish (Parkes) and ATCA observations \citep{waugh05}.  The ASKAP \HI\ recessional velocities follow the optical $v=cz$ definition and the $S_{\rm{HI}}$ listed in this table includes the flux correction factor described by \citet{for20}.}
\label{HItab}
\begin{scriptsize}
\begin{tabular}{lccc}
\hline
\hline
Property & Parkes & ATCA & ASKAP (this work) \\
(1)    & (2)         & (3)   & (4)   \\
\hline
\multicolumn{4}{c}{C1 properties}\\
\hline
RA &   03:39:09.6 & 03:39:12.5 &03:39:12\\
Dec & $-$22:23:47 & $-22$:23:10 &$-22$:23:22 \\
$V_{\rm{50}}$ (\kms)& 1876$\pm4$ & 1879 $\pm4$ & 1879 $\pm2$ \\
$W_{\rm{50}}$ (\kms)& 41$\pm4$ & 28 $\pm2$ & 23 $\pm2$  \\
Beam & 15.5\arcmin\ &113\arcsec\ $\times$ 297\arcsec\ & 30\arcsec\ $\times$ 30\arcsec\ \\
RMS (mJy~beam$^{-1}$)& 7.0  & 6.5  & $\approx$2.5  \\
$S_{\rm{HI}}$ (Jy~\kms) & 3.0 $\pm0.6$  & 0.8   &  2.28 $\pm 0.18$\ \\
$M_{\rm{HI}}$ (\msun)&$3.1 (\pm0.6) \times 10^8$ &$0.8\times 10^8$ & $2.37 (\pm0.19) \times 10^8$ \\
$D_{\rm{HI}}$ (kpc) & ---& --- & 6.4 $\pm 1.1$\\
\hline
\multicolumn{4}{c}{C2 properties}\\
\hline
RA &   03:37:31.7 & 03:37:24.7 &03:37:23 \\
Dec &$-$23:59:33 & $-$23:58:25 & $-23$:57:54\\
$V_{\rm{50}}$ (\kms)& 1467 $\pm7$  & 1471 $\pm3$ &1469 $\pm2$ \\
$W_{\rm{50}}$ (\kms)& 41$\pm7$ & 17 $\pm6$ & 16 $\pm2$ \\
Beam &15.5\arcmin\ &113\arcsec\ $\times$ 279\arcsec\ &30\arcsec\ $\times$ 30\arcsec\ \\
RMS (mJy~beam$^{-1}$)& 7.0  & 6.5  & 2.4  \\
$S_{\rm{HI}}$ (Jy~\kms)& 2.5 $\pm0.6$ &  0.9  & 1.35 $\pm 0.16$  \\
$M_{\rm{HI}}$ (\msun) &$2.6 (\pm0.6) \times 10^8$ & $0.9 \times 10^8$ & $1.41 (\pm0.16) \times 10^8$ \\
$D_{\rm{HI}}$ (kpc) & ---& --- & 7.0 $\pm 1.2$\\
\hline                     
\end{tabular}
\end{scriptsize}
\end{table}

%%%%%%%%%%%%%%%%%%%%%%%%%%%%%%%%%%%%%%%%%%%%%%%%%%%%5
\section{Results}
\label{sectionresults}
\begin{figure*}
\begin{tabular}{cc}
\includegraphics[scale=.23]{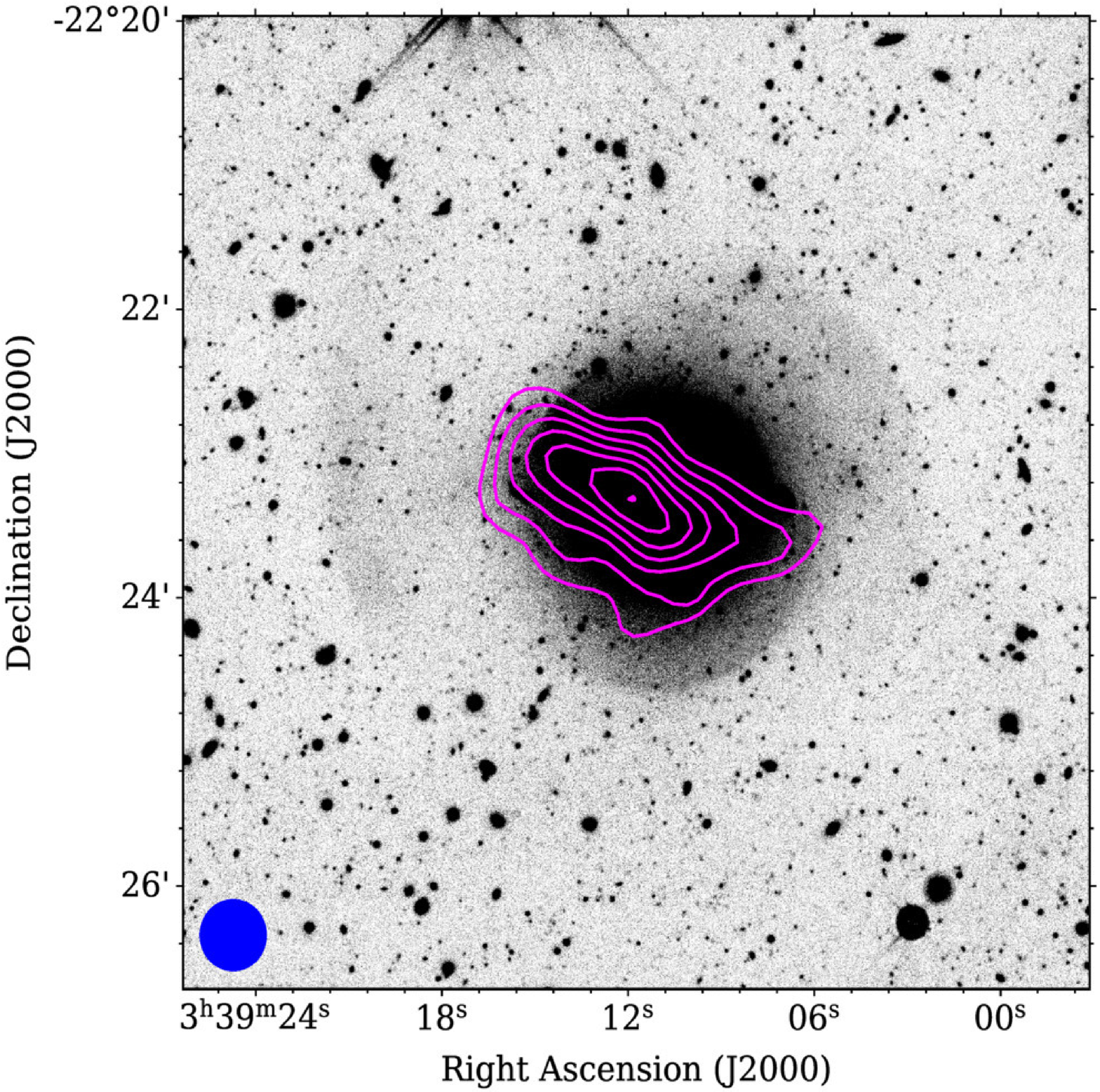} &\includegraphics[scale=.23]{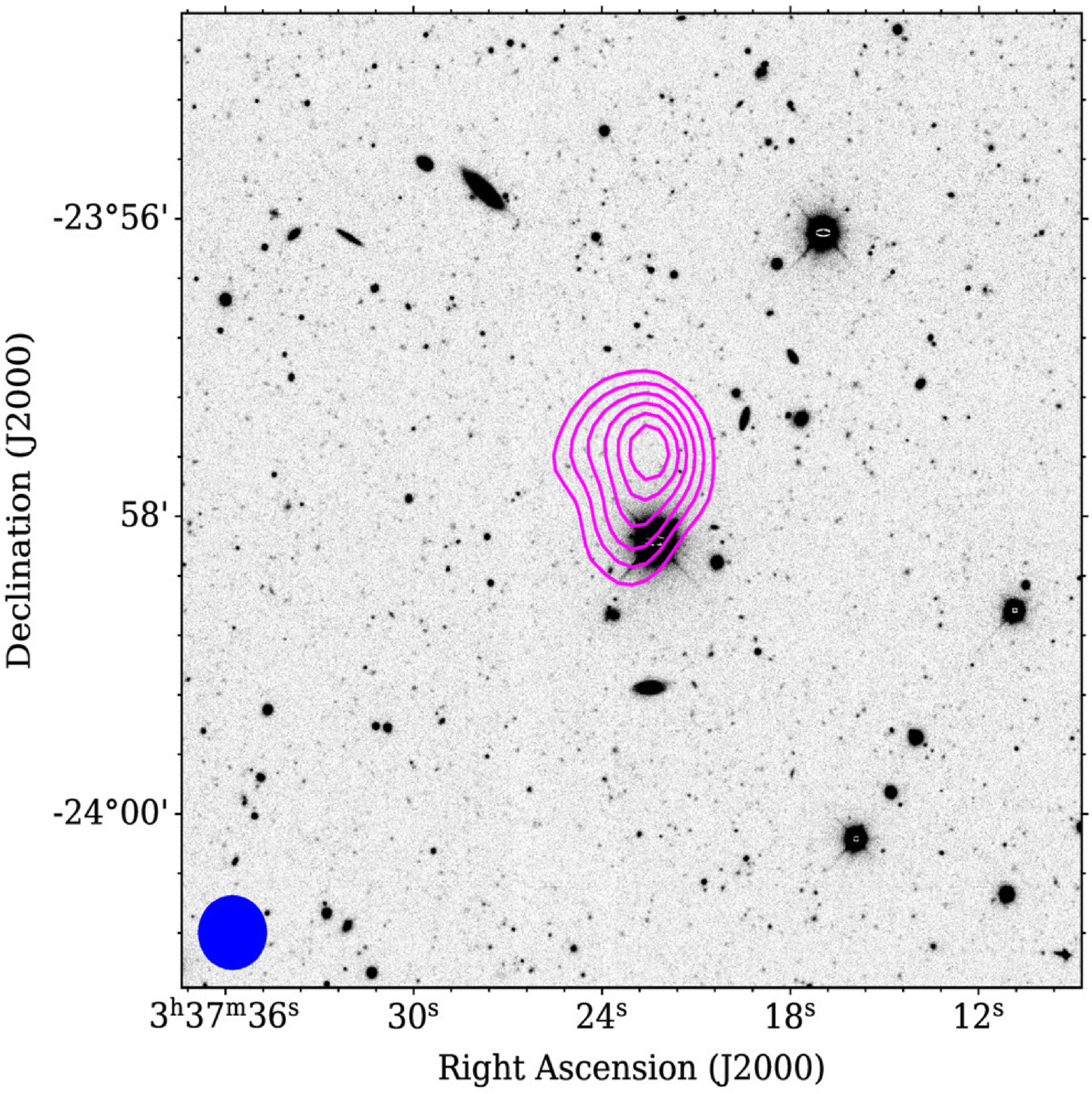} \\
\includegraphics[scale=.3,angle=270]{./fig2c.ps} & \includegraphics[scale=.3,angle=270]{./fig2d.ps} \\
\includegraphics[scale=.32]{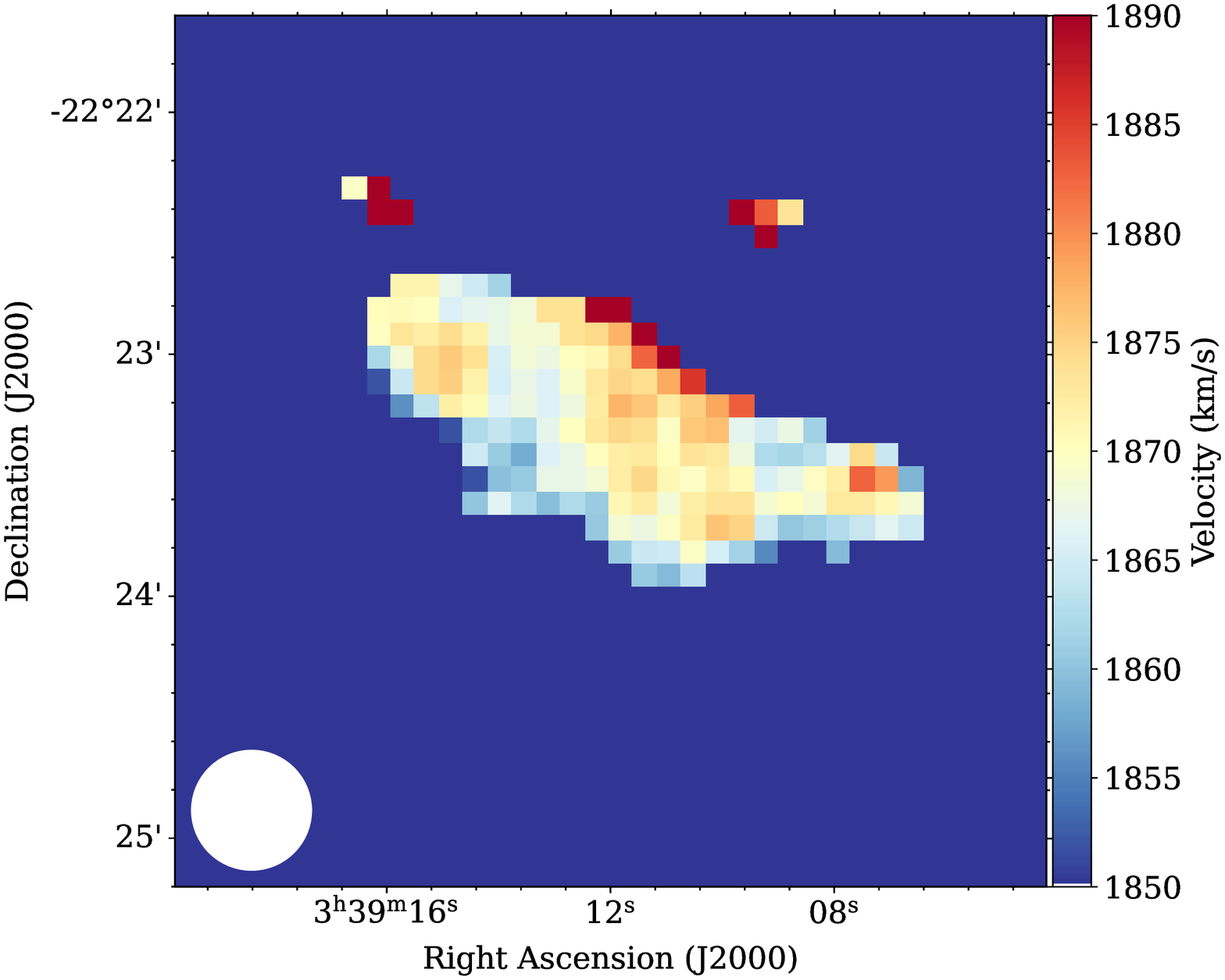}  &\includegraphics[scale=.32]{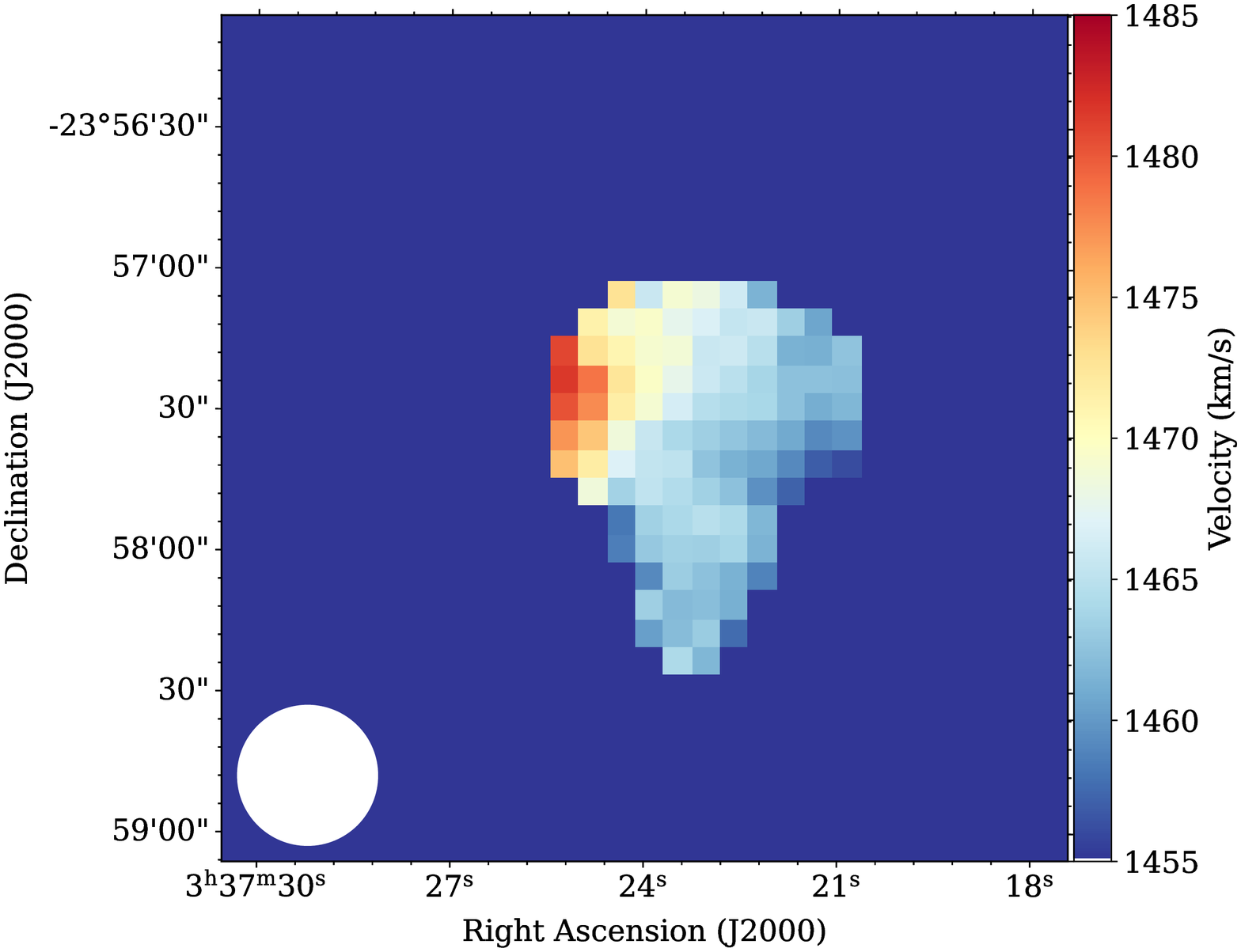} \\
\end{tabular}
\caption{Top row: Moment zero \HI\ column density maps of C1 (left column) and C2 (right column) overlaid on a deep optical image stack from DR8 and DR9 of the DECam Legacy Survey  \citep{dey19}. Stellar shells from the background ETG behind C1 is shown by the deep DR9 imaging stack.  We overlaid C2's \HI\ contours onto the DR8 optical image in order to show the location of the intervening foreground star.  This foreground star has been subtracted in the DR9 optical images.  The lowest \HI\ contour levels for C1 and C2 begins at $1.1 \times 10^{20}$~atoms~cm$^{-2}$  and subsequent contour levels are at increments of $0.5 \times 10^{20}$~atoms~cm$^{-2}$. The blue circle in the bottom-left of each panel in the top row represents the 30-arcsecond ASKAP synthesised beam.  The middle row shows the \HI\ spectra of both C1 (left) and C2 (right). The vertical lines mark the maximum velocity width of the \HI\ profiles. The bottom row shows the three-arcminute by three-arcminute moment 1 maps of both C1 (left) and C2 (right).  The \HI\ emission below the \HI\ column density of $1.1 \times 10^{20}$~atoms~cm$^{-2}$ has been masked from these moment one maps.   }
\label{cloud}
\end{figure*}

\subsection{The \HI\ properties of C1 and C2}
\label{hiclouds}
 
  Our ASKAP position and velocity centres are flux-weighted centroids estimated by SoFiA.  Table~\ref{HItab} presents the \HI\ properties of C1 and C2 as parametrized by the SoFIA source finder \citep[version 2.0.1; ][]{sofia,westmeier20}, in addition to \HI\ properties measured by  previous single-dish (Parkes Basketweave) and synthesis (ATCA) observations.  Relative to the Basketweave detections, these new ASKAP observations find positional offsets of 37~arcsec and 2.6~arcmin for C1 and C2, respectively. Such positional offsets are unsurprising since we are comparing between Parkes observations with a 15.5~arcmin beam and ASKAP observations with a synthesized beam of 30~arcsec.  In addition, ASKAP may resolve out the \HI\ emission that exists on larger angular scales and consequently contribute towards an observed positional offset between the two sets of observations.  When comparing between two sets of interferometric observations, we find smaller positional offsets between previous ATCA and ASKAP observations of 12\arcsec\ and 41\arcsec\ for C1 and C2.  The velocity centres and linewidths of both C1 and C2 from the recent ASKAP observations are consistent with those found by the Parkes Basketweave survey and the ATCA observations. Additionally, the ASKAP and ATCA velocity widths for C1 and C2 are in good agreement,
but lower than the Basketweave values. Figure~\ref{cloud} shows the \HI\ moment zero maps of C1 and C2 overlaid as magenta contours over deep optical images, in addition to their \HI\ spectra. The lowest column density contour displayed corresponds to the 3-sigma column density level across the \HI\ linewidth of each source.

%\begin{table*}
%\begin{center}
%\caption{\HI\ properties of C1 and C2 measured by the SoFiA source finder.}
%\label{HItab}

%\begin{tabular}{llccccccc}
%\hline
%\hline
%Source & WALLABY ID &  RA (J2000) & Dec (J2000) & $v_{\rm{HI}}$ &  $S_{\rm{HI}}$ & $RMS$ &  $W_{\rm{50}}$ & $M_{\rm{HI}}$ \\
%(1)    & (2)         & (3)   & (4)         & (5)          & (6)  & (7) & (8) &(9)\\
%\hline
%C1 & WALLABY J033911-222322 & 03:39:12 & $-22$:23:22 &1879  &1.90 ($\pm 0.03$)  & 2.13&  23 &$2.33$ ($\pm 0.03$) $\times10^8$\\
%C2 & WALLABY J033723-235753 &03:37:23 & $-23$:57:54 & 1469  &1.13 ($\pm 0.02$) & 2.41& 16 &  $1.38$ ($\pm 0.03$) $\times10^8$\\
%\hline
%WALLABY J033723-235745 & 0.753 ($\pm 0.002$)  & 5.90 mJy& 1468 ($\pm 0$) & 19 &  $9.20 (\pm0.03) \times10^7$\\
%WALLABY J033911-222322 & 1.386 ($\pm 0.086$) & 5.80 mJy & 1876 ($\pm 2$) & 24 & $ 1.69 (\pm0.11) \times10^8$\\
%\hline                     
%\end{tabular}    
%\end{center}
%\raggedright Column (1): Source name in this paper; Column (2): WALLABY identification \citep{for20}; Column (3): Right Ascension (J2000) of the flux-weighted centroid position; Column (4): Declination (J2000) of the flux-weighted centroid position; Column (5): \HI\ central velocity in \kms; Column (6): Integrated flux in Jy~kms$^{-1}$; column (7): rms values in mJy~beam$^{-1}$; column (8): $W_{50}$ in \kms; Column (9): Estimated \HI\ mass in solar masses assuming a distance of 22.8~Mpc. Uncertainty values are statistical uncertainties estimated by SoFiA \citep{serra15}.
%\end{table*}

 Including the $\sim20$ percent $S_{\rm{HI}}$ correction factor reported by \citet{for20} for the ASKAP pre-pilot observations, we find that the ASKAP-derived $S_{\rm{HI}}$ for C1 is consistent with that found from the Parkes single-dish observations.   On the other hand, the ASKAP observations detect only 54~percent of $S_{\rm{HI}}$ that was  found in C2 from Parkes \citep{waugh05}.  The significant \HI\ flux deficit in the ASKAP measurements of C2 is consistent with, and likely related to the positional offset noted earlier, when comparing these ASKAP interferometric observations to Parkes single-dish observations.

 % In comparison to previous HIPASS detection of C1's $S_{\rm{HI}}$, we find that these ASKAP observations only measured approximately 58~per-cent of that from HIPASS.  Furthermore a comparison to the Basketweave survey, finds that ASKAP detected 0.63 and 0.45  of the \HI\ emission ($S_{\rm{HI}}$) reported by \citet{waugh05} for both C1 and C2.  As discussed previously in Section~2.2, missing diffuse \HI\ at larger angular scales may be one explanation for the observed \HI\ deficit in these ASKAP observations relative to previous single dish observations.  
%On the other hand, we recover a factor of 2.4 and 1.3 more $S_{\rm{HI}}$ for C1 and C2, respectively, relative to previous interferometric observations of both \HI\ sources using ATCA by \citet{waugh05}.  As these previous ATCA observations had synthesised beams that range from 2 to 5~arcmin \citep{waugh05}, the missing \HI\ from the ASKAP observations is likely to be due to emission at angular scales that are closer to that of the Parkes beam, as both C1 and C2 are unresolved point sources in HIPASS and the Basketweave Survey. 

The \HI\ velocity widths listed in Table~\ref{HItab} and the spectra in Figure~\ref{cloud} show that both C1 and C2 have narrow \HI\ linewidths.  Qualitatively, it is unclear from the \HI\ velocity maps (Figure~\ref{cloud}) whether C1 and C2 are in stable rotation. To further investigate the possibility for rotating motions, we attempt to model the \HI\ kinematics of C1 and C2.  First we obtained a single Gaussian fit of the velocity field via a Bayesian Markov Chain Monte Carlo profile decomposition algorithm  \citep[BAYGAUD; ][]{oh19} which provided the input parameters into an automated two dimenstion (2D) Bayesian tilted-ring fitting tool, 2DBAT \citep{oh18}.  We do not find evidence for a clear rotation pattern in C1.  We are able to fit a rotating model to C2 using two 2DBAT setups.  The first setup had a constant position angle and inclination while fitting a cubic spline for the rotation velocity.  In the second setup,  only the inclination was held constant and cubic splines are fitted for the rotation velocity and the position angle.  Both setups resulted in rotating models with a maximum rotation velocity of $\sim35$~\kms.  We note that our results are significantly affected by beam smearing as C2 is only marginally resolved with only approximately 2 beams across its rotating axis.  
 
In marginally resolved observations such as those for C2, it is more accurate to fit three dimensional (3D) kinematic models that take beam smearing effects into account, so we modelled the 3D kinematics of C2 using TiRiFiC \citep{jozsa07,jozsa21}.  We created two models for C2 where we assume that C2 is a finite-thickness disk  with a flat rotation curve, a constant sech$^2$ scaleheight and a constant velocity dispersion.  One model assumed that we are observing C2 at a constant inclination of 90~degrees (edge-on) and the second model allowed the inclination to vary (for a flat rotation curve) along with the surface brightness.  For the edge-on model, we find a rotation amplitude of  $7.2\pm0.5\,\mathrm{km}\,\mathrm{s}^{-1}$ and a velocity dispersion of $8.1\pm 0.3\,\mathrm{km}\,\mathrm{s}^{-1}$.  
For the second model, we find a rotation of  $22.7\pm5.7\,\mathrm{km}\,\mathrm{s}^{-1}$ at an inclination of $19.6\pm4.9\,\mathrm{deg}$ and a velocity dispersion of $8.4\pm 0.2\,\mathrm{km}\,\mathrm{s}^{-1}$.  It is clear from these 3D kinematic models that C2 is likely to be rotating and both models fit our observations equally well due to the low angular resolution of our ASKAP observations (see Figure~\ref{pvel} in Appendix~\ref{posvel}).  We note that it may also be possible for C2 to be rotating and observed face-on. However, we are not able to narrow down the inclinations and position angles of C2 (and consequently our models) from our ASKAP \HI\ observations alone. Hence, our 3D kinematic models suggest that C2 is likely to have a true rotation velocity between $7.2\pm0.5\,\mathrm{km}\,\mathrm{s}^{-1}$ and $22.7\pm5.7\,\mathrm{km}\,\mathrm{s}^{-1}$.

%\begin{figure*}
%\begin{tabular}{cc}
%\includegraphics[scale=.3,angle=0]{lucycloud.eps} &\includegraphics[scale=.3,angle=0]{othercloud.eps}\\
%\end{tabular}
%\caption{Placeholder Footprint B maps of potential \HI\ clouds. Note that while Lucy (left) looks like a non-rotating cloud, there is a chance that J0337 (right) may be LSB behind a Galactic star.}
%\label{cloud}
%\end{figure*}

\subsection{Optical search for stellar counterparts to C1 and C2}
\label{dark}
 To investigate the presence of stellar components that could be associated with C1 and C2, we use
the deep optical imaging from the publicly-available Dark Energy Camera Legacy Survey Data Release 9
\citep[DECaLS DR9; ][]{dey19}.
In addition to the improvement in surface brightness over previous data releases, DECaLS DR9 provides images
where bright Galactic foreground stars have been modelled and subtracted.  This is especially useful for our
investigation of C2's potential optical counterpart.

\subsubsection{C1}
%Both `dark' sources have no detectable optical stellar counterparts at the detection limit of the archival Dark Energy Survey's DR1 obervations (see Figure~\ref{cloud}).$\mu_{lim,g}$ = 29.4 mag arcsec$^{-2}$ and $\mu_{lim,r}$ = 29.1 mag arcsec$^{-2}$. This is approximately 3 mag arcsec$^{-2}$ deeper than the SDSS data used in the identification of the optical counterparts of the HI-rich ultra-diffuse galaxies by Leisman + 17, ensuring that objects similar to these would be clearly detected in our optical data. I

The spectroscopic observations using the 2.3~m telescope at the Siding Springs Observatory (described in Section~2.3) find that NGC~1403 has a
recessional velocity of 4173~($\pm 135$) ~\kms\  --- confirming that C1 is a foreground object that is associated
with  the Eridanus group, and not NGC~1403 as previously identified (Table~\ref{HItab}). This recent measurement confirms previous spectroscopic observations and is 
consistent with results from peculiar velocity modelling by Cosmicflows-3 \citep{tully16,tully19,pomarede20}. Velocity and density field reconstructions from Cosmicflows-3 find that
 C1 and the rest of the Eridanus group are travelling towards the Great Attractor; while NGC~1403 resides in a weak filament in the Sculptor void and is moving towards the Perseus-Pisces cluster.  Figure~\ref{cflow} in Appendix~\ref{cosflows3} illustrates the density field and velocity vectors of C1 (marked with a '1')
 and NGC~1403 (marked with a '2') determined by  Cosmicflows-3.

At the location of C1, additional processing of the DECaLS DR9 images enables us to reach a surface brightness limit of $\mu_{lim,g} = 29.4$~mag~arcsec$^{-2}$ and $\mu_{lim,r} = 29.1$~mag~arcsec$^{-2}$ in the $g$- and $r-$ bands, respectively. A detailed description for the method used to estimate the surface brightness limit can be found in Appendix~A of \citet{roman20}.  As  C1 is projected in front of NGC~1403, we aim to differentiate between the optical component of NGC~1403 and any optical component associated with C1 by subtracting a simple  model of NGC~1403 from the DR9 image.   While NGC~1403 is an ETG with a complex set of stellar shells, we assume that the central stellar body of NGC~1403 is consistent with a stellar density profile that is described by a Sersic model with an index of $n=4$. 

\begin{figure}
\begin{center}
\includegraphics[scale=.2,angle=0]{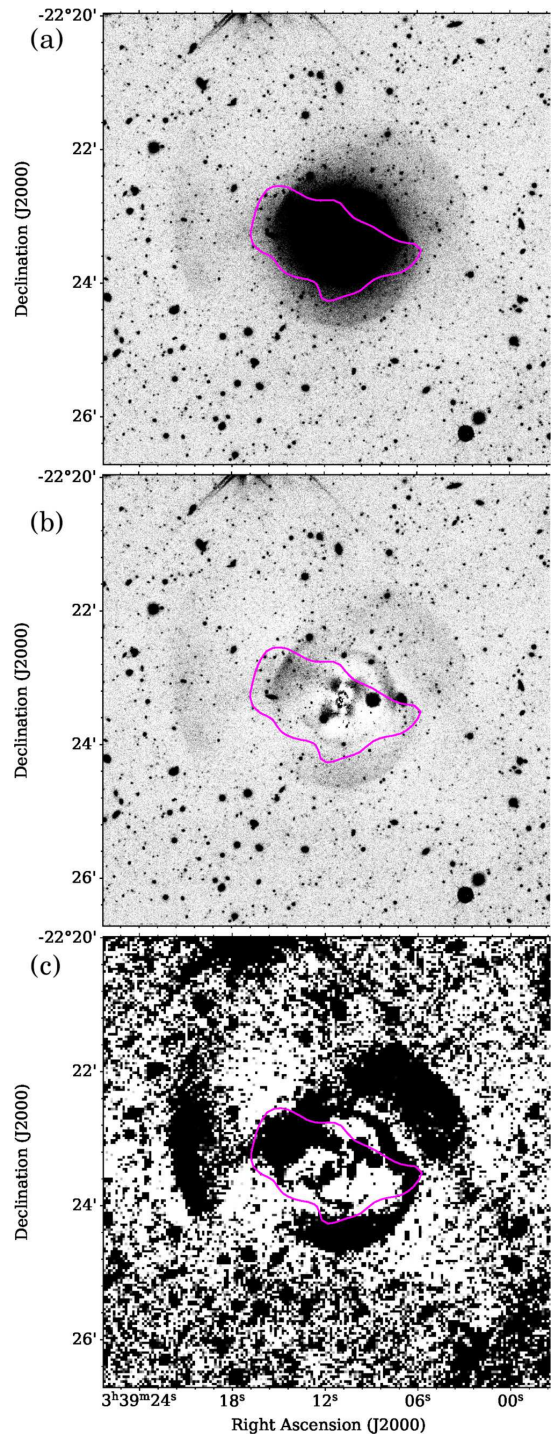}
\end{center}
\caption{DECaLS DR9 $g$-band optical images of C1.  The lowest column density  \HI\ moment zero contour at $1.1 \times 10^{20}$~cm$^{-2}$ is shown as a magenta contour.  Panels (a) and (b) show the optical images of NGC~1403's before and after an early-type galaxy model is subtracted, respectively.  Panel c is a rebinned residual image where the residual image (Panel b) is rebinned to a coarser pixel resolution to further enhance the surface brightness limit. All stellar structure correlate to NGC~1403's large stellar shell complexes and a distinct stellar component from an additional low surface brightness galaxy is absent.}
\label{c1dr9}
\end{figure}

The deep optical imaging from DECaLS DR9 reveal a set of complex stellar shells that are affiliated with  NGC~1403 (Figure~\ref{c1dr9}a).  Figure~\ref{c1dr9}b shows the residual optical image of NGC~1403 after the subtraction of our simple ETG model.  To further enhance the signal-to-noise of any diffuse low surface brightness structures remaining in Figure~\ref{c1dr9}b, we rebinned Figure~\ref{c1dr9}b to a coarser pixel scale (Figure~\ref{c1dr9}c).  Figures~\ref{c1dr9}b and c show that the diffuse stellar components are connected and not independent of NGC~1403's stellar shells.  Therefore we do not observe any distinct stellar component that could be associated with C1 and distinguishable from NGC~1403's stellar structures.  For greater context, the surface brightness limit attained from our DR9 images is approximately 3~mag~arcsec$^{-2}$ deeper than the SDSS data used in the identification of the optical counterparts for the HI-rich ultra-diffuse galaxies by \citet{leisman17}, ensuring that similarly low surface brightness objects can be detected in our optical images.

\subsubsection{C2}
\begin{figure}
\begin{center}
\includegraphics[scale=.2,angle=0]{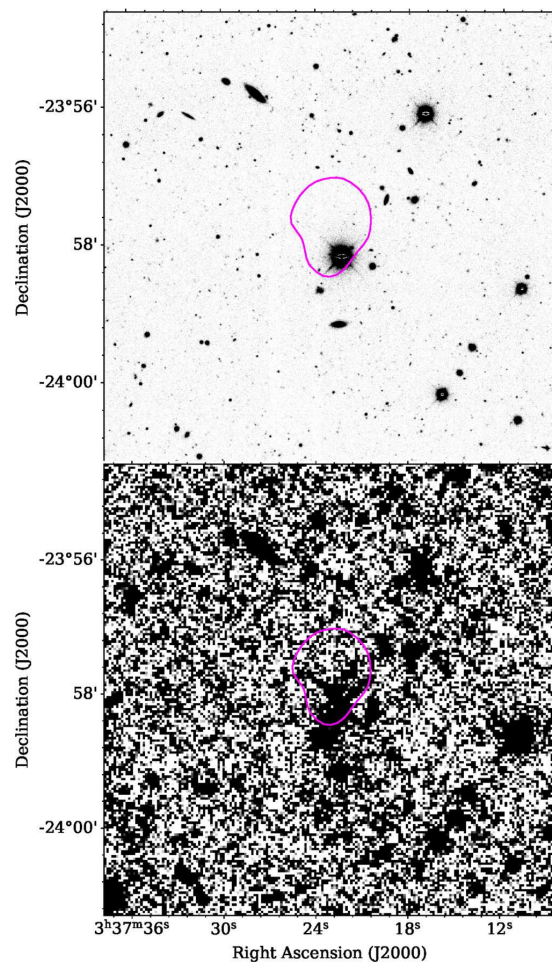}
\end{center}
\caption{DECaLS DR8 and DR9 $g$-band optical images of C2.  The lowest column density  \HI\ moment zero contour at $1.1 \times 10^{20}$~cm$^{-2}$ is shown as a magenta contour. The top panel shows the DECaLS DR8 image before the subtraction of the foreground star. The bottom panel shows the DR9 image where the foreground star has been subtracted and the entire image has been rebinned to a courser resolution to further enhance the surface brightness of any faint stellar feature.  The main structure observed in the bottom panel relates to residuals from the incomplete removal of the foreground star and no additional stellar structure is discernable that could be related to C2.}
\label{c2dr9}
\end{figure}

Similar to C1, we investigate the presence of a stellar counterpart to C2 using the deep optical observations from DECaLS DR9 where the bright foreground star that is coincident with C2's position has been modelled and subtracted.  The top panel in Figure~\ref{c2dr9} shows the DECaLS DR8 $g$-band image before the intervening foreground star is subtracted. The bottom panel of  Figure~\ref{c2dr9} shows the DECaLS DR9 $g$-band image where the star has been subtracted.  The image has been rebinned to a courser pixel resolution to further enhance the surface brightness sensitivity. The dominant emission appears to have arisen from the residuals of the stellar subtraction.  We do not find a low surface brightness component that could be related to C2.

\section{Discussion: \HI\ debris from past interactions or dark galaxy candidates?}
\label{sectiondisc}

In this section, we investigate the possible origin and nature of these two \HI\ sources.  Specifically, we explore the possibility that: 1)  C1 and C2 originated from tidal interaction(s);  and 2) C1 and C2 are primordial dark (or almost-dark) galaxies through investigating potential analogues from cosmological simulations.  We conclude this section by comparing the \HI\ properties of C1 and C2 to other classes of \HI-dark (and low surface brightness) objects.  We also discuss the implications of our results.
%, as well as to that of optically-dark \HI\ cloud that was previously found by \citet{kilborn06} in an environment that is similar to that of the Eridanus subcluster.

%Both C1 and C2 are massive \HI\ sources (within the virial radius of the NGC~1395 subcluster) with no stellar components detected down to the surface brightness limits of the Dark Energy Survey (DES) DR1's observations (Figure~\ref{cloud}).  A search for the diffuse stellar counterpart to both C1 and C2 was also undertaken using archival ultraviolet and near-infrared using the GALEX and WISE public archives, also returned no stellar counterparts. Are C1 and C2 dark (very low surface brightness) galaxies or remnant \HI\ clouds from previous interactions with nearby galaxies?  

%To further investigate the possible origin and nature of these two \HI\ sources, we we search for such systems in the Illustris TNG cosmological simulations.

\subsection{Investigating a tidal origin}
The two optically-dark WALLABY \HI\ sources flanking the massive early-type galaxy, NGC~1395, may have originated from past galaxy interactions. Examples of \HI\ tidal remnants through galaxy group interactions have been reported in previous ASKAP WALLABY early-science observations \citep[e.g.\ ][]{serra15,lee19} as well as MeerKAT observations \citep[e.g.\ ][]{namumba21}.  

A recent study of  Fornax A (NGC~1316), another nearby ETG with comparable properties (stellar mass, shells, accretion history) to NGC~1395  suggested that  \HI\ streams extending 70 -- 150~kpc away from the galaxy can result from multiple  10:1 merger events \citep{schweizer80,iodice17,serra19}.  The projected separations between C1 with NGC~1395 and C2 with NGC~1395 are at 238~kpc and 363~kpc, respectively --- beyond the 135~kpc radius of NGC~1395's  X-ray halo as well as its stellar shells \citep{brough06,tal09,escudero17}.  Following the results of \citet{kravtsov13}, we estimate NGC~1395's  $r_{\rm{200}}$ (the radius at the density which is 200 times that of the critical density of the Universe) to be approximately 354~kpc assuming an optical effective radius of 48~arcsec \citep{lauberts89}. %
At these projected distances from NGC~1395, both C1 and C2 are not consistent with the projected distance of 15 effective radii that are typically observed in optically-selected samples of tidal dwarf galaxies \citep[TDG; ][]{kaviraj12}, which can be classed as almost-dark galaxies due to their low stellar surface brightnesses.  These younger and less-evolved TDGs are still embedded within the tidal debris of their parent galaxies \citep[e.g.\ ][]{lelli15}.  
On the other hand, older and more evolved TDGs could be located further way from their parent galaxies \citep{bournaud06}.   Hence, we are not able to rule out the possibility that C1 and C2 are more evolved TDGs, from the projected separations alone.  Regardless of projected separations, the young diffuse stellar populations within a TDG will likely fade into an almost-dark or dark galaxy within 2~Gyr, due to the aging of the stellar population \citep{roman21}.

To differentiate between the possibility that C1 and C2 are candidate tidal remnants as opposed to candidate TDGs, we can compare our observed $W_{\rm{50}}$ for C1 and C2, to the simple expectation of circular velocity ($v_{\rm{G}}$) from a self-gravitating gas cloud via  $v_{\rm{G}} = \sqrt{G \times 1.33 M_{\rm{HI}} / R_{\rm{HI}}}$ where $G$ is the gravitational constant, $M_{\rm{HI}}$ and $R_{\rm{HI}}$ are the \HI\ mass and \HI\ radius, respectively.  A factor of 1.33 is included to account for the contribution of Helium.    We estimate $v_{\rm{G}}$ to be 20.6~\kms\ and 15.2~\kms\ for C1 and C2, respectively.   Even though we only recover 54~percent of the integrated \HI\ flux of C2, the $v_{\rm{G}}$ values estimated for both C1 and C2 under the assumption of self-gravity, are  consistent with the observed $W_{\rm{50}}$ (Table~\ref{HItab}).  This comparison suggests that C1 and C2 are consistent with the circular velocities of simple self-gravitating systems, thereby favouring the hypothesis that C1 and C2 are TDG candidates over that of tidal remnants.

To further constrain the potential for more local interactions, we examine the nearest neighbouring galaxies to both C1 and C2.  While the estimation of $r_{\rm{200}}$ for massive neighbouring galaxies is useful for constraining the possibility for gravitational interactions occuring between two galaxies, it is less useful for higher galaxy density environments where the halo potential is complicated by overlapping $r_{\rm{200}}$ from individual subhaloes.   This is indeed the case for C1.  Therefore,  we use a tidal index, $\Theta$ \citep{karachentsev13,pearson16},  to quantify the tidal influence between a neighbouring massive galaxy ($M_{*} > 10^{10}$~M$_{\odot}$) relative to C1 and C2.  Following \citet{pearson16}, $\Theta$ is defined to be
\begin{equation}
  \Theta = \log_{10}\left[ \frac{M_*}{10^{11} {\rm M}_\odot} \left( \frac{D_{\rm proj}}{\rm Mpc} \right)^{-3} \right]
\end{equation}  
where $M_{*}$ is the stellar mass and $D_{\rm{proj}}$ is the projected distance between C1 (or C2) and their nearest massive galaxy neighbour.  Following \citet{pearson16}, isolated environments are defined by $\Theta <0$, intermediate and high density environments are traced by $0< \Theta <1.5$ and $\Theta >1.5$, respectively.

\begin{figure*}
\includegraphics[scale=.5]{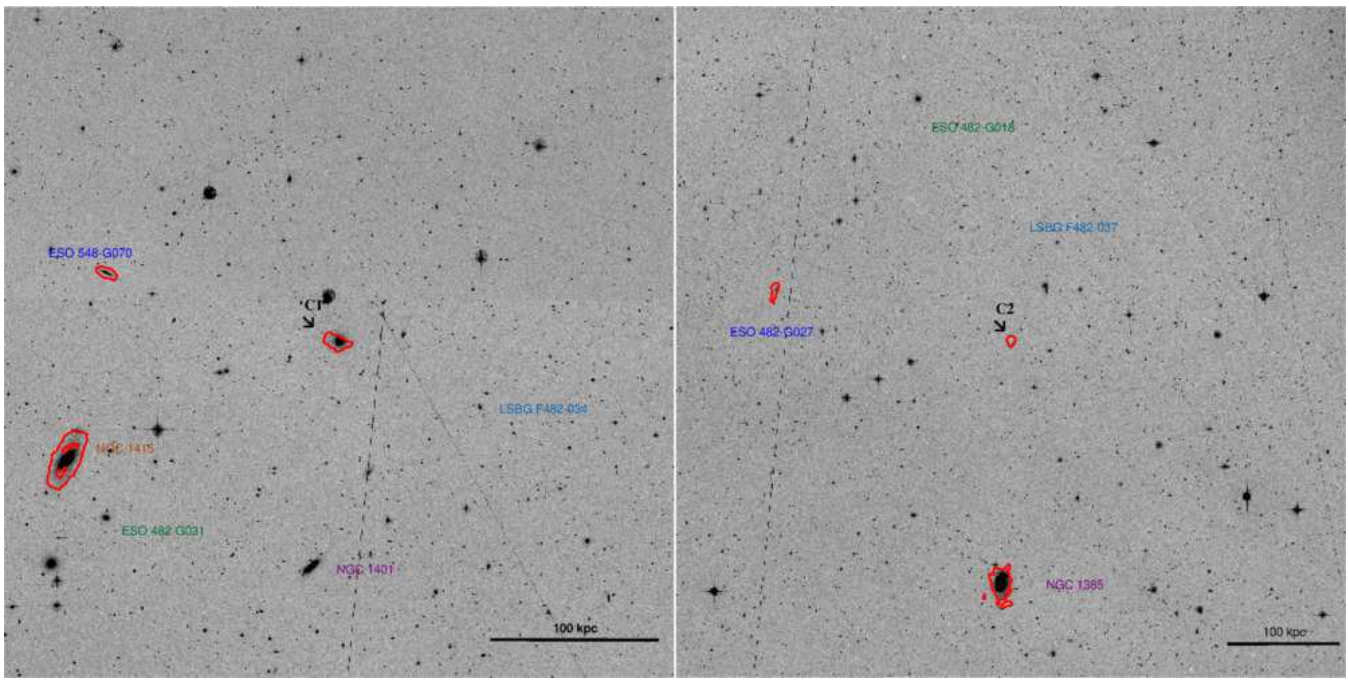}
\caption{Map of C1 and C2, showing their nearest projected galaxy neighbours within 500~\kms.  The WALLABY \HI\ moment zero emission at the level of $1.1 \times 10^{20}$~cm$^{-2}$ is outlined by the red contour, overlaid on a DSS blue background image centred on each of the dark WALLABY \HI\ sources.   A physical scale bar is provided in the bottom right corner of each panel.  It should be noted that galaxies that are unlabelled reside in a more distant galaxy group at greater recessional velocities than C1 and C2. Also, the grey dashed line should be ignored as it is an artefact. Left panel: a one-degree by one-degreed map centred on C1.   Right panel: a 1.5~degree by 1.5~degree map centred on C2. North and East are to the top and left of the page, respectively.}
\label{lucyfriends}
\end{figure*}

The nearest massive galaxies (where  log~$M_{*}$~$\geq$~10.0~\msun) to C1 and C2 (in projection) are NGC~1401 and LSBG~F482-037, respectively. Table~\ref{c1tidaltab} and Table~\ref{c2tidaltab} list the tidal indices determined for C1 and C2 with respect to their nearest projected massive  neighbours. Figure~\ref{lucyfriends} shows the galaxies neighbouring both C1 and C2. We find that $\Theta =2.0$ for C1 and NGC~1401.  Using the stellar mass measured by \citet{for20} for LSBG~F482-037 of $1.26 \times 10^{10}$~M$_{\odot}$, we find that $\Theta =2.3$ for C2.
These tidal indices suggests that C1 and C2 may be under high tidal influence. However, the lack of \HI\ in both NGC~1401 and LSBG~F482-037 suggests that C1 and C2 are less likely to originate from  NGC~1401 and LSBG~F482-037, respectively.

The nearest massive galaxies to C1 and C2 (in projection) for which \HI\ is detected are ESO~548-G070 and NGC~1385, respectively.  Located at larger projected distances, the resulting $\Theta$ values suggest that C1 and C2 may be under intermediate strength tidal influence.  Since the tidal indices are only an approximate tracer of the tidal fields within which C1 and C2 reside, we cannot rule out the possibility that C1 and C2 have had a tidal origin.  This is especially the case for C2 where our ASKAP observations appear to be missing 46~percent of the total integrated \HI\ that was measured from  the Parkes Basketweave survey \citep{waugh05}.  Deeper \HI\ observations will likely recover the true extent and further constrain the origin of C2.

%Furthermore, C1 has two other massive \HI-rich neighbours (ESO~548-G070 and NGC~1415) at slightly larger projected distances than that of NGC~1401.  Therefore the tidal index may not be an accurate tracer of the tidal field within which C1 resides.  

\begin{table}
\caption{Tidal indices ($\Theta$)  determined between C1 and its nearest massive neighbours.}
\label{c1tidaltab}
\begin{center}
\begin{tabular}{lccc}
\hline
\hline
Neighbour & log~$M_{*}$& Projected distance & $\Theta$\\
  & (\msun) & (kpc) & (log$_{\rm{10}}$ [\msun~Mpc$^{-3}$])\\
(1)    & (2)         & (3) & (4)   \\
\hline
NGC 1395 & 10.9 & 238  & 1.8 \\
NGC 1401 & 10.3 & 127  & 2.0 \\
NGC 1415 & 10.3 & 163  &  1.7 \\
ESO 482-G031 & 10.1&  159 & 1.5 \\
\hline                     
\end{tabular}    
\end{center}
\end{table}

\begin{table}
\caption{Tidal indices ($\Theta$) determined between C2 and its nearest massive neighbours.}
\label{c2tidaltab}
\begin{center}
\begin{tabular}{lccc}
\hline
\hline
Neighbour &  log~$M_{*}$&Projected distance & $\Theta$\\
  & (\msun) & (kpc) & (log$_{\rm{10}}$ [\msun~Mpc$^{-3}$])\\
(1)    & (2)         & (3)   & (4) \\
\hline
NGC 1395 & 10.9 & 363  & 1.2 \\
NGC 1385 & 10.0 & 197  & 1.2 \\
ESO 482-G018 &10.3 & 219 & 1.3 \\
LSBG F482-037 &10.1 & 90 & 2.3 \\
\hline                     
\end{tabular}    
\end{center}
\end{table}

%We find that \HI\ emission is not detected in every neighbouring galaxy.  Both dark \HI\ sources reside within approximately 215~kpc (in projection) of two neighbouring galaxies with detected \HI.

\begin{figure}
\includegraphics[scale=.7]{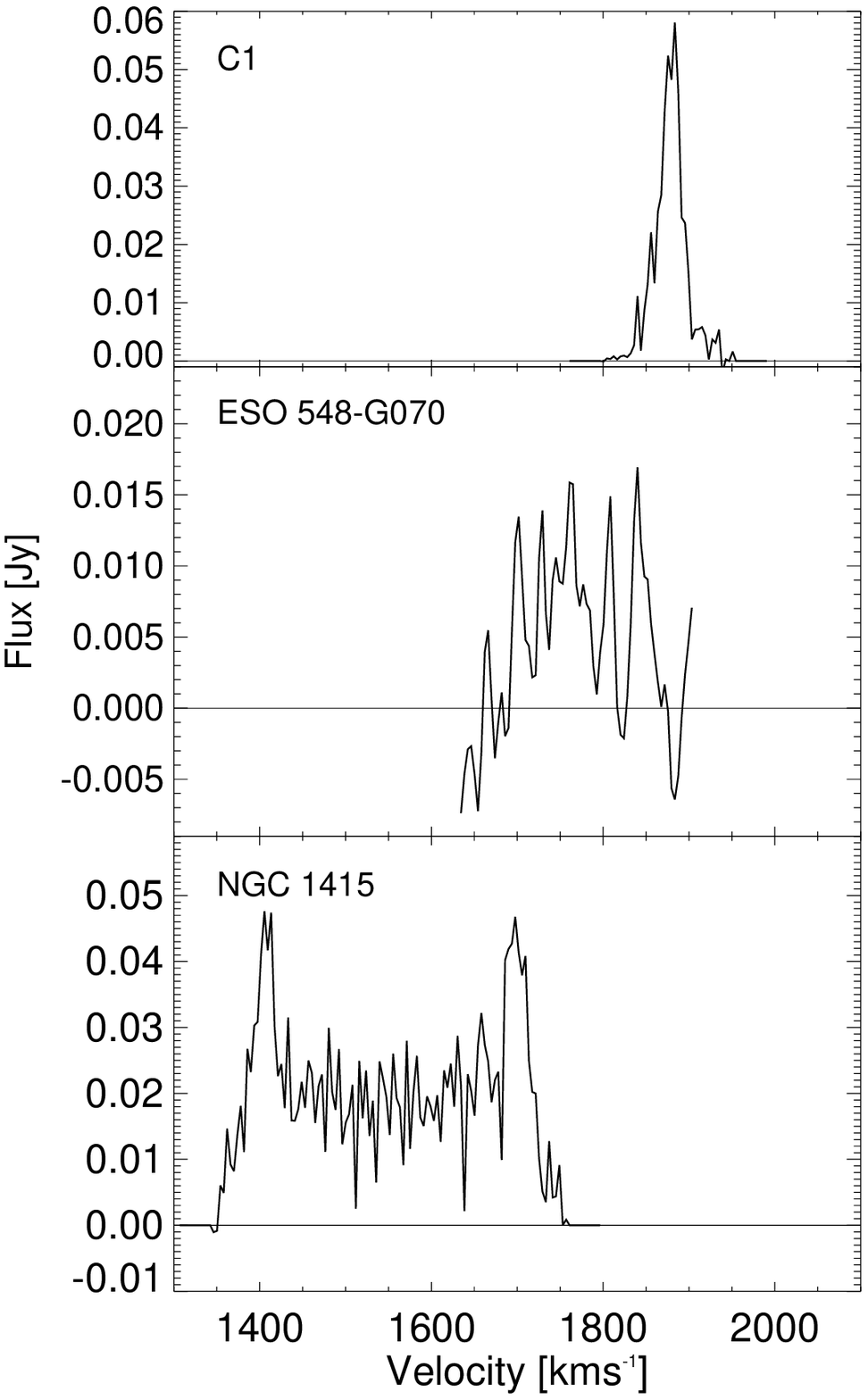}
\caption{A comparison of C1's \HI\ spectrum (top panel) to that of its neighbouring \HI-rich galaxies identified by SOFIA. It should be noted that ESO~548-G070 and NGC~1415 have WALLABY source names, WALLABY~J034040-221711 and WALLABY~J03406-223350 \citep{for20}.   }
\label{lucyneigbourspec}
\end{figure}

The \HI\ spectra of NGC~1415 and ESO~548-G070 are at lower recessional velocities than C1 (see Figure~\ref{lucyneigbourspec}), and we do not observe any signatures within the \HI\ morphology or \HI\ kinematics that would suggest any recent interactions.   While we cannot completely rule out any interaction between all three \HI\ sources in this field, the lack of any \HI\ tail or extended structure does not provide strong support for interactions between the sources. The ASKAP \HI\ centre position of C1 and \HI\ mass are similar to that found from the Basketweave survey. The \HI\ observations of C1 also do not appear to indicate that the \HI\ is rotating.  Follow-up observations with higher spectral resolution may provide further constraints on the kinematics and possibly the origin of C1.

\begin{figure}
\includegraphics[scale=.7]{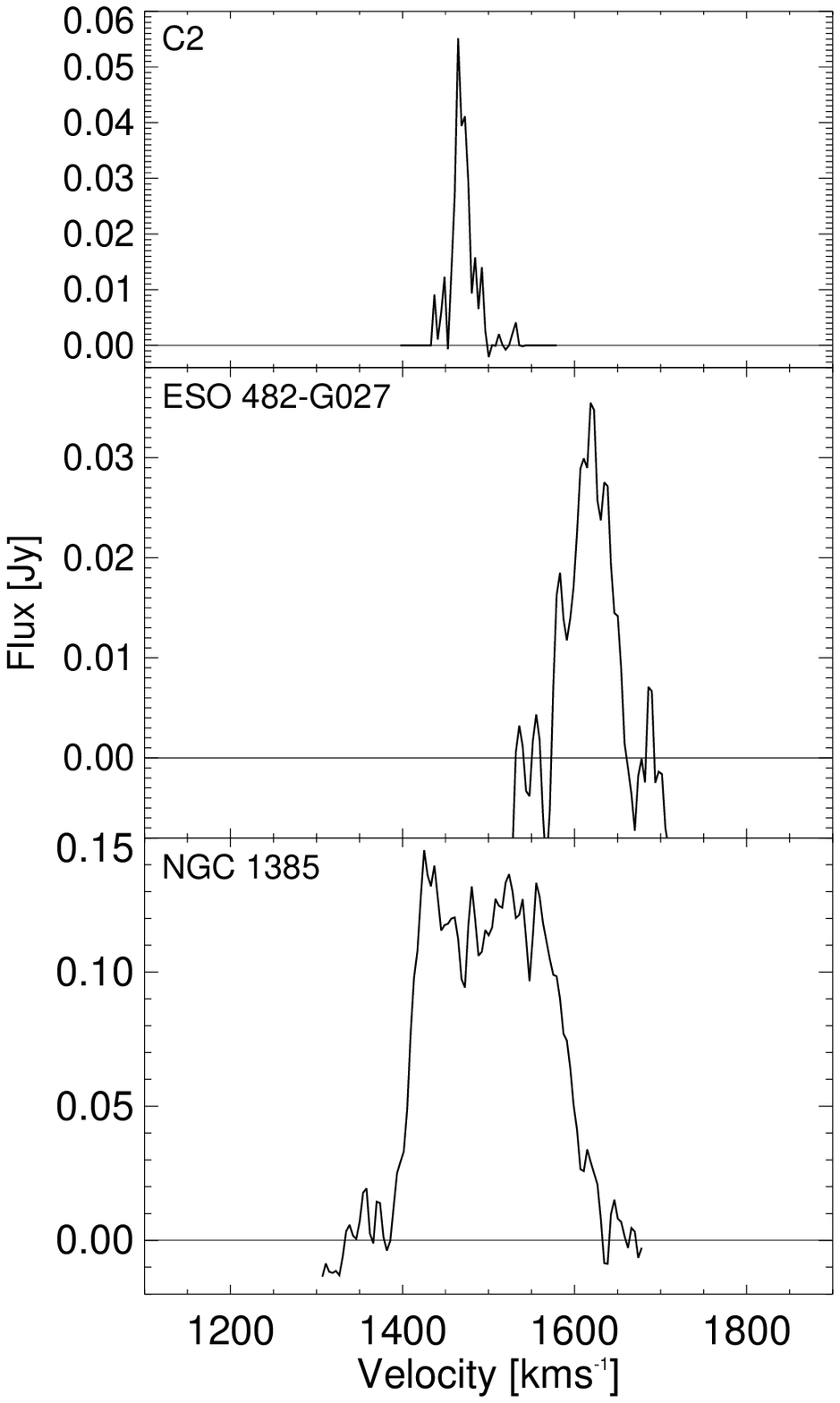}
\caption{A comparison of C2's  \HI\ spectrum (top panel) to that of its neighbouring \HI-rich galaxies identified by SOFIA. It should be noted that ESO~482-G027 and NGC~1385 have WALLABY source names, WALLABY~J033941-235054 and WALLABY~J033728-243010 \citep{for20}.  }
\label{friendneigbourspec}
\end{figure}

C2 resides in a region south-west of NGC~1395 with four neighbouring galaxies.  C2 is 32.4\arcmin\ ($\approx 214$~kpc in projection) north of NGC~1385 (the closest \HI-rich neighbour).  While we do not  find any \HI\ emission extending between NGC~1385 and C2, both sources have consistent recessional velocities (see Figure~\ref{friendneigbourspec}). Also, the \HI\ morphology of C2  and the lack of \HI\ in LSBG~F482-037 suggests that if C2 was a product of a past interaction, it is more likely to have originated from NGC~1385.
  This idea is supported by the fact that  Parkes observations of C2 found the \HI\ position centre to be 2.6\arcmin\ South-East of the ASKAP position centre and that the total \HI\ emission from previous single-dish observations is nearly a factor of two greater than that observed by ASKAP. Hence, deeper follow-up \HI\ observations may reveal more diffuse \HI\ emission that could extend southwards towards NGC~1385.

%C2 is also only 14.7~\arcmin\ south-east of a low surface brightness galaxy (LSBG~F482-037) but \HI\ has not been detected in this galaxy.  The \HI\ morphology of C2 and the lack of \HI\ in LSBG~F482-037 suggests that if C2 was a product of a past interaction, it is more likely to have originated from NGC~1385.  This idea is supported by the fact that previous Parkes observations of C2 found the \HI\ position centre to be 2.6\arcmin\ South-East of the ASKAP position centre and that the total \HI\ emission from previous single-dish observations is greater by a factor of 2.2 to that observed by ASKAP.   While the ASKAP observations does not find any \HI\ emission extending between NGC~1385 and C2, both sources have consistent radial velocities (see Figure~\ref{friendneigbourspec}).  Hence, it is possible that C2 possesses a significant amount of diffuse \HI\ that extends southwards towards NGC~1385.  

In summary, the tidal indices estimated for both C1 and C2 suggest that we cannot rule out the tidal influences of local galaxy neighbours, and hence a possible tidal origin for these clouds.  Despite the higher tidal indices estimated between C1 and C2 relative to their closest projected massive galaxy neighbours, we argue that \HI\ from these closest projected neighbours are unlikely to be origins of both C1 and C2  due to the lack of \HI\ in both nearest neighbours. The tidal indices estimated for C1 and C2 relative to their closest projected \HI-rich neighbours  suggest an intermediate-level of tidal influence. As the ASKAP \HI\ observations of C1 fully recover the total integrated \HI\ previously found from the Basketweave survey, C1 is a good candidate for an evolved TDG.  Similarly, the observed \HI\ velocity widths of C1 and C2 are also consistent with the approximate circular velocities of self-gravitating gas clouds, which suggests that C1 and C2 are more likely to be TDG candidates than young tidal remnants.   As we do not observe any tidal tails in C2 at the  measured 3$\sigma$ column density sensitivity of $1.1 \times 10^{20}$~cm$^{-2}$, future analysis of C2 would benefit from the feathering of single-dish observations with these ASKAP synthesis observations in order to probe the larger scale diffuse emission at lower column densities. Such future work would allow us to study the \HI\ emission and structure at both arcsecond and arcminute scales.

\subsection{Investigating a primordial origin}

To explore the possible evolutionary history of C1 and C2 as primordial dark galaxy candidates (DGCs), we searched for sources with similar properties to C1 and C2 in the TNG100 cosmological simulations. The difference between DGCs discussed in this section and that of the almost-dark TDGs from the previous subsection is that the DGCs in this subsection formed in the early Universe, and not from tidal interactions in the last few billion years.  Also, the primordial DGCs discussed in this section reside in their own dark matter halo prior to entry into the cluster or group environment at later times.

The TNG simulations were conducted with {\sc{arepo}} \citep{springel10}, and follow $\Lambda$CDM with cosmological parameters from the Planck Collaboration \citep[$\Omega_{\rm{m}}=0.3089$, $\Omega_{\rm{b}}=0.0486$, $h=0.6774$, $\sigma_{\rm{8}}=0.8159$; ][]{ade16}. Prescriptions that describe the physical processes of baryons, including below the resolution limit, accounting for gas cooling, star formation, growth of massive black holes, magnetic fields and feedback from both stars and active galactic nuclei can be found in \citet{weinberger17} and \citet{pillepich18}. TNG100 has a box length of $75/h \! \simeq \! 110$~Mpc and a baryonic mass resolution of  $\sim \! 1.4 \! \times \! 10^6\, {\rm{M}}_{\odot}$.  Gas cells were post-processed to compute their atomic- and molecular- Hydrogen masses \citep{diemer18,stevens19a}.  We use the following criteria for selecting analogue DGCs for C1 and C2 in TNG100:
\begin{itemize}
\item satellite galaxies with stellar masses below $10^{7.5}$~\msun. By TNG100 definition, such a stellar mass limit does imply an optically-dark galaxy.
\item \HI\ masses between $10^{8.0}$--$10^{8.5}$~\msun\
\item subhalo dark matter fractions above 5~per~cent (to avoid numerical artifacts)  
\item host haloes of mass $10^{13} \! \leq \! M_{\rm{200}}/$\msun$ \! \leq \! 10^{14.5}$, whose central galaxies have neutral gas-to-stellar fractions that are less than 0.01. This halo mass range is consistent with that of the Eridanus group as well as the total halo mass of the eventual merger of all three subclusters within the Eridanus Cluster as per \citet{brough06}. 
\end{itemize}

We find three DGCs following this set of criteria. The three DGCs in TNG100 (hereafter known as TNG-DGC)  reside within host haloes with log$_{10} \! \left(M_{\rm{200}}/{\rm{M}}_{\odot} \right)=$  14.33, 13.80, and 13.19.  All three TNG-DGC (hereafter labelled by their subhalo IDs in the TNG public database -- 14081, 246257 and 88856) have dark matter fractions ranging from 97~percent to 98~percent.

%\begin{figure}
%\includegraphics[scale=.3]{../SIMS/SM_DM_dist.eps}
%\caption{Halo and stellar mass of the 3 TNG analogues.   }
%\label{tng-mass}
%\end{figure}

%\begin{figure*}
%\includegraphics[scale=.3]{../SIMS/Infall.eps}
%\caption{Trajectory of 3 TNG analogues of C1 and C2.   }
%\label{tng-infall}
%\end{figure*}

%\begin{figure*}
%\includegraphics[scale=.35]{../SIMS/tng_ionisedgasdens.eps}
%\caption{Evolution history of the three TNG-DGC (shown individually in three columns). Top row: Mass as a function of redshift and look-back time.   The \HI\ prescriptions from \citet{gnedin11} (GK11), \citet{gnedin14} (GD14), \citet{krumholz13} (K13) are shown as cyan dot-dashed lines, cyan long dashed lines and cyan short dashed lines, respectively. The grey shaded region marks the point in time when these modelled galaxies are in-falling into a cluster environment.      Bottom row: Temperature and density of the ISM gas as a function of redshift and look-back time within the three TNG-DGC.     }
%\label{tng-comps}
%\end{figure*}
\begin{figure*}
\includegraphics[scale=.35]{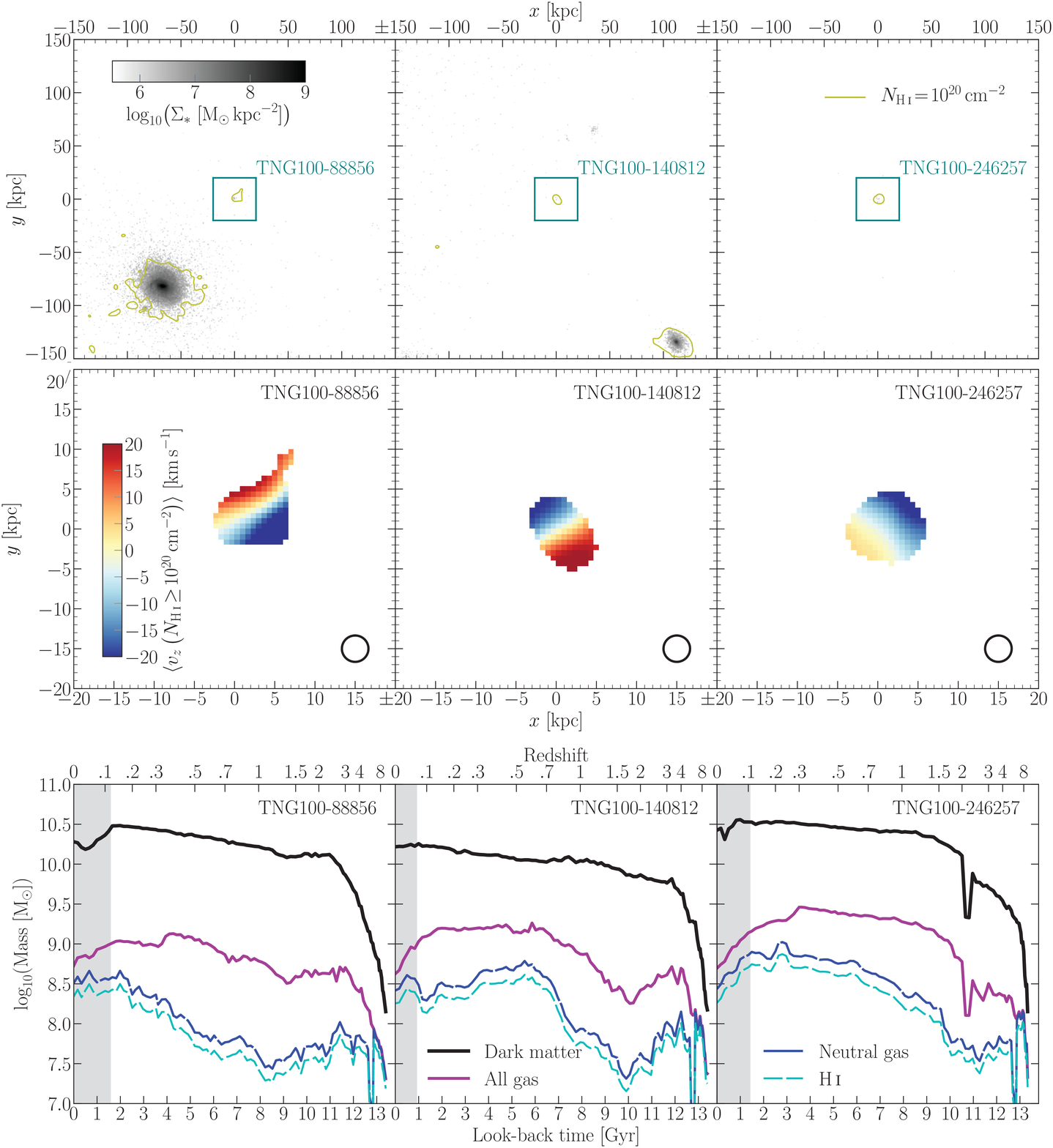}
\caption{Images and evolutionary history of the three TNG-DGC (shown individually in three columns).  Top row of panels: Images of each TNG-DGC and their immediate surroundings at $z\!=\!0$.  Greyscale is used to indicate stellar surface density, including any material within the host halo of the DGC (integrated along the line of sight).  Contours show an \HI\ column density of $10^{20}$~atoms~cm$^{-2}$.  The $(x,y) \! = \! (0,0)$ position refers to the DGC's centre of potential, as given in the TNG public database \citep{nelson19}. Middle row of panels: \HI-mass-weighted mean-velocity (moment 1) maps in the line of sight ($z$-direction) of each TNG-DGC.  A minimum \HI\ column density of $10^{20}$~atoms~cm$^{-2}$ is imposed for the maps, where only gas bound to the TNG {\sc subfind} subhalo of interest is considered.  The boundary of each panel matches the teal square in the top row of panels.  For the purposes of visualisation, we have approximated each gas cell as a sphere (subsequently collapsed into two dimensions), maintaining the cell's actual volume in the simulation (see the {\tt build\_gas\_image\_array()} function at {\url{https://github.com/arhstevens/Dirty-AstroPy/blob/master/galprops/galplot.py}}). The circle in the bottom right corner of each panel represents the FWHM of the Gaussian kernel (FWHM=3.3~kpc) with which the sources were convolved, in order for these models to be consistent with the observed WALLABY resolution.  Bottom row of panels: Mass growth history of the TNG-DGCs.  Each DGC is tracked back through the TNG {\sc SubLink} merger trees \citep{rodrigues-gomez15}, where all mass associated with the relevant {\sc subfind} subhalo at each snapshot is summed.  `All gas' means exactly that: ionised, neutral, hydrogen, and all other chemical species summed.  `Neutral gas' accounts for all atomic and molecular chemical species.  The grey shaded region mark when each TNG-DGC became a satellite.  To account for the reduced information available in TNG `mini snapshots', we follow appendix B of \citet{stevens21} to calculate the neutral and \HI\ mass at each snapshot for these panels. }
\label{tng-comps}
\end{figure*}

Figure~\ref{tng-comps} presents the evolution history for each of the three TNG-DGCs.  The top  row of Figure~\ref{tng-comps} shows the local environment in which each TNG-DGC resides at $z\!=\!0$ within a projected distance of $\sim$150~kpc.  The middle row of Figure~\ref{tng-comps} shows the Moment 1 (\HI-mass-weighted mean-velocity) maps for each TNG-DGC projected along the line of sight.  While our ASKAP \HI\ observations do not show clear rotation in C1 or C2,  the TNG-DGC kinematics very clearly show rotating bulk motion. The bottom row of Figure~\ref{tng-comps} shows the mass growth history of the TNG-DGCs.  Assuming that the three TNG-DGC are bona fide analogues of C1 and C2, we can infer that C1 and C2 would also have built up most of their dark matter mass within the first 2~Gyr of formation and remained in a relatively uneventful evolutionary history.  We note that the \HI\ masses derived from TNG used three independent prescriptions \citep{gnedin11,gnedin14,krumholz13} yielded results which  are consistent (with each other) for all three TNG-DGCs.% and in general follow the mass evolution for total gas within each modelled galaxies. The exception to this is
%In the past 1~Gyr, TNG-DGC 140812  appears to have decreased its total gas mass but have increased its \HI\ mass.  On the other hand, both TNG-DGC 246257 and 88856 have lost total gas and \HI\ gas as they fall into their respective clusters.

These TNG-DGCs suggest that \HI-rich, DGCs within clusters could have formed at early times and remained dark due to a fairly uneventful evolutionary history whereby the  cluster in-fall (in the past 1 to 2~Gyr) has only affected its total gas mass but that gas losses have not significantly impacted the star formation rate of each of these galaxies. Hence, such DGCs will likely remain dark unless they are on orbital trajectories that lead to stronger interactions with other galaxies or their local environment.

Table~\ref{tngprop} lists the  \HI\ properties of the three TNG-DGCs. We emphasise that the \HI\ properties provided are not mock ASKAP properties but are TNG model properties. 
The three TNG-DGCs have properties  close to the stellar mass and halo mass resolution limit of TNG100 but with sufficient particle resolution for a statistically-accurate determination of global galaxy properties.  A distinct difference between the three TNG-DGCs compared to C1 and C2 is that the TNG-DGCs are clearly rotating and have greater \HI\ velocity widths (W50; Table~\ref{tngprop}), likely due to its significant dark matter fraction. On the one hand, the  TNG-DGCs' greater \HI\ velocity widths could be due to the TNG-DGCs having had more  time to build up their angular momentum, relative to that of C1 and C2. This is consistent with the suggestion that disk galaxies with high spins would be Toomre stable against star formation \citep{jimenez20}.

On the other hand, the  greater \HI\ velocity widths of the TNG-DGCs may be associated with the larger sizes (angular extents) that are typically found in TNG galaxies (see Section~\ref{compare}), a systematic feature of the TNG100 simulations that has been previously studied in \citet{stevens19b}.  As such, we are not able to conclude with absolute certainty that C1 and C2 have a primordial origin even though the TNG-DGCs are the closest model analogues from TNG100 to C1 and C2.

\begin{table}
\caption{Dark matter (DM) and \HI\ properties of the TNG-DGCs.}
\label{tngprop}
\begin{footnotesize}
\begin{center}
\begin{tabular}{lccc}
\hline
\hline
Property & T100-88856& T100-140812 & T100-246257\\
(1)    & (2)         & (3) & (4)   \\
\hline
\HI\ W50 (\kms) & 65.2 &65.3 & 41.0 \\
\HI\ radius (kpc) & 4.1 &3.7 & 5.0 \\
\HI\ mass (\msun) & $1.0\times10^8$ & $1.1\times10^8$ &$1.3\times10^8$\\
DM mass (\msun) &$2.0 \times10^{10}$&$1.8\times10^{10}$  &$2.8\times10^{10}$ \\
\hline                     
\end{tabular}    
\end{center}
\end{footnotesize}
\end{table}

\subsection{Comparison to other known low surface brightness objects}
\label{compare}

In addition to the \HI\ fluxes and moment maps, our new ASKAP \HI\ observations are at sufficient surface brightness sensitivity and angular resolution that we are able to compare both C1 and C2's \HI\ sizes and \HI\ masses to other known `dark' (and almost-dark) low surface brightness sources.  Such a comparison may provide further insights towards our understanding of such a class of sources that while currently rare, are increasing in numbers as the next generation \HI\ surveys get underway.
We determine the \HI\ diameters ($D_{\rm{HI}}$) of C1 and C2 out to a gas surface density of 1~M$_{\odot}$~pc$^{-2}$ to be 6.4~kpc and 7.0~kpc, respectively (Table~\ref{HItab}).

\begin{figure}
\includegraphics[scale=.75]{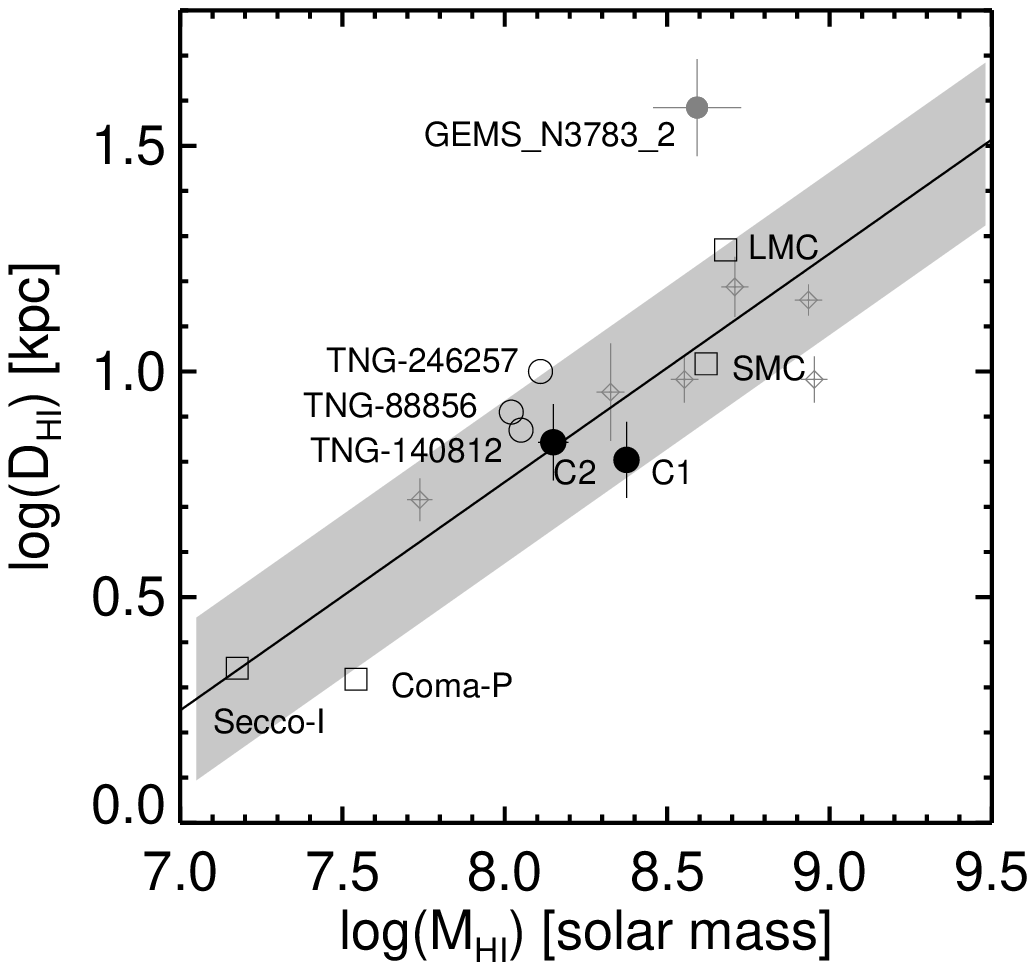}
\caption{The two dark \HI\ sources (C1 and C2; solid black circle) are consistent with the \HI\ size--mass relationship of low-mass galaxies in the Local Universe (solid line and grey-shaded 3$\sigma$ scatter) from \citet{wang16}.  The \HI\ size--mass properties of the Small and Large Magellanic Clouds \citep{staveley98,staveley03}, Secco-I \citep{sand17} and Coma-P \citep{brunker19} are shown as open squares for comparison purposes. The TNG-DGCs are represented by black open circles and the rotationally-supported TDGs from \citet{lelli15} are shown as small grey open diamonds.  The previously-discovered optically-dark \HI-rich source GEMS\_N3783\_2 \citep{kilborn06} is represented by a large grey solid circle. }
\label{szmass}
\end{figure}

The  \HI\ size--mass relationship is typically observed in rotationally supported galaxies, and has been attributed to the small range of \HI\ gas densities within galaxies \citep{wang16,stevens19b}. 
Figure~\ref{szmass} compares the \HI\ sizes and masses of C1 and C2 to the TNG-DGCs described in the previous subsection, as well as other low surface brightness,  almost-dark low mass galaxies such as the Magellanic Clouds, Secco-I and Coma-P \citep{brunker19}, a sample of rotating TDGs from \citet{lelli15}, as well as a previously-reported optically-dark \HI\ source (GEMS\_N3783\_2) that resides in the NGC~3783 galaxy group \citep{kilborn06}. 
In general, the three TNG-DGCs have \HI-sizes and masses that are consistent with the \HI\ size--mass relationship from \citet{wang16}.  The TNG-DGCs have similar \HI\ masses to C1 and C2 but larger diameters than those of C1 and C2.  %We note that the larger \HI\ extents of the TNG-DGCs are consistent with the understanding from \citet{stevens19b} that the \HI\ sizes derived from the TNG100 model galaxies are marginally more extended on average than the \HI\ sizes that are empirically measured for galaxies in the Local Universe \citep[e.g.\ ][]{wang16}.

Figure~\ref{szmass} shows that both C1 and C2 are: 1) consistent with the \HI\ size-mass relationship that is found for low-mass and low surface brightness galaxies in the Local Universe, including rotationally-supported TDGs; 2)  are only a factor of a few less massive in \HI\ than both the Small and Large Magellanic Clouds; and 3) are factors of a few more massive than the `almost-dark' galaxies, Coma-P \citep{brunker19} or Secco-I \citep{sand17}. The consistency of C1's and C2's \HI\ properties with that of the \HI\ size-mass relationship may be suggestive of rotational support at some level even if our current ASKAP \HI\ observations only show support for rotation in C2.
On the other hand, the previously-discovered \HI-rich optically dark source, GEMS\_N3783\_2 is less consistent with the \HI\ size--mass relationship. Perhaps GEMS\_N3783\_2 is at an earlier stage of evolution? Previous active searches for \HI-rich optically dark sources in Local Volume galaxy groups have also revealed dark massive \HI\ sources to be very rare and concluded that these massive \HI\ clouds are likely to originate from strong galaxy interactions \citep{chynoweth11}.  From Figure~\ref{szmass} alone, it appears that C1 and C2 have \HI\ sizes and masses that are consistent with other rotating, low-mass galaxies whether or not C1 and C2 have a past tidal or primordial origin.

%Unlike the Magellanic Clouds, it may be possible that both C1 and C2 have stellar surface brightnesses that are more comparable to that of Coma-P \citep{janowiecki15}.

In terms of the stellar content, we are not able to identify any stellar component for either C1 or C2 (Section~\ref{dark}).   Assuming that the observed rotation in C2 implies that C2 is stable and is consistent with a stable disk star formation model described by \citet{wong16}, we could expect the star formation rate (SFR) of C2 to range between 0.01 and 0.03~M$_{\odot}$~year$^{-1}$.   For comparison, the more massive (than C1 and C2) Small Magellanic Cloud (SMC) has a SFR of 0.05~M$_{\odot}$~year$^{-1}$ \citep[e.g.\ ][]{wilke04}.  The  stellar mass of the SMC is $1.5 \times 10^9$~\msun\ \citep{wilke04} and  has sufficient stellar surface brightness to be detectable at the distance of the Eridanus group. A simple back-of-the-envelope estimation suggests that C2 could build up approximately 10$^7$~\msun\ of stars if it were able to sustain this SFR (of 0.01 -- 0.03 M$_{\odot}$~yr$^{-1}$) for 1~Gyr or more.    Even so,  C2 is still more than two orders of magnitude less massive (and likely at lower stellar surface density) than the SMC.  Furthermore, we know from \citet{roman21} that the optical surface brightness will further diminish as the stellar population ages.   Hence, it is unsurprising that we are not able to detect a significant stellar population within C2.

\subsection{Implications}
\label{sectionimpl}
The upcoming WALLABY survey provides an important step forward in the study of \HI\ galaxy evolution as it overcomes the angular-resolution limitation from previous single-dish large-area \HI\ surveys such as HIPASS \citep{barnes01} and ALFALFA \citep{giovanelli05}, through synthesis imaging.  On the other hand, interferometric observations trades angular resolution for surface brightness sensitivity. Furthermore, the intrinsic lack of zero-spacing information in the visibility ($uv$-) plane translates to a potential for `resolving out' emission that exists on large angular scales.  So far, WALLABY early-science commissioning and pilot observations have successfully recovered the total \HI\ emission from a diverse set of \HI\ observations of galaxies in environments ranging from the field to groups and clusters \citep{serra15,reynolds19,lee19,elagali19,kleiner19,for19,wang21}.

The ASKAP pre-pilot observations presented in this paper targets a galaxy group that is in an active state of cluster infall, and galaxies within this group are understood to be experiencing active `pre-processing' \citep{for20,murugeshan20}.  Therefore, the probability of \HI\ existing at a large range of densities and angular extents is higher than in non-interacting galaxies that reside in less dynamic environments.  While our ASKAP synthesis observations did not fully recover the total integrated \HI\ emission for C2 from previous Parkes observations, we note that we have recovered more \HI\ emission than previous ATCA synthesis observations, even though the ATCA observations had a larger synthesised beam than ASKAP. This demonstrates the superior performance in {\em{both}} surface brightness sensitivity and angular resolution that we can expect from the upcoming WALLABY survey using ASKAP.  On the other hand, deeper \HI\ observations to better surface brightness sensitivities are required to map the full extent of diffuse tidal structures \citep[e.g.\ ][]{namumba21}.

The two dark \HI\ sources reported in this paper are potential candidates for an extreme class of objects that would not have been found through targeted surveys of optically-bright galaxies.  While the \HI\ kinematics of the TNG-DGC do not match those of C1 and C2, we are not able to completely rule out a primordial origin without more sensitive and higher resolution follow-up \HI\ observations and modelling.  
We recover the same total integrated \HI\ flux for C1 as previously measured from single-dish observations, but we only recover 54~percent of the integrated \HI\ flux for C2.

If both C1 and C2 have a tidal origin and are TDG candidates, then C1 and C2 may be at different stages of TDG evolution. Based on the complete recovery of the total \HI\ emission of C1 and the agreement in position between our ASKAP observations and those from Parkes,  we can argue for C1 to be at a more advanced stage of TDG evolution relative to that of C2.  On the other hand, the observed rotation in C2's \HI\ emission does favour C2 to be at a more stable self-gravitating state than C1 where rotation is absent.  The possibility also remains that both C1 and C2 to have originated through different formation and evolutionary processes.  Such heterogeneity in formation pathways is consistent with recent findings that low surface brightness galaxies (or ultra-diffuse galaxies) have multiple formation and evolutionary pathways \citep[e.g.\ ][]{yozin15,papastergis17,roman17,bennet18,ferre18,janowiecki19,sales20}.  As \HI-rich low surface brightness galaxies represent a subset and possibly a younger demographic of the low surface brightness Universe, such samples are useful in the study of star formation at low densities and efficiencies --- an area of active and current research \citep[e.g.\ ][]{mcgaugh17,thilker19}.

Further analysis of C1 and C2, including the \HI\ emission at larger angular scales, as well as more sophisticated methods for the identification of the possible stellar components will be required to verify whether these sources are remnants of past tidal interactions or bona fide extreme examples of almost-dark galaxies.  We note that it is non-trivial to identify the stellar components of both C1 and C2 from deep optical or near-infrared imaging as there is a chance alignment along the line of sight of C1 and C2 with a background early-type galaxy and a foreground star, respectively.

\section{Conclusions}
\label{sectionsum}
We present new ASKAP WALLABY pre-pilot \HI\ observations of two dark \HI\ clouds projected within 363~kpc of NGC~1395 in the Eridanus group, WALLABY~J033911-222322~(C1) and WALLABY~J033723-235745~(C2).  Previous association of C1 to NGC~1403 is incorrect, and no detectable stellar components could be attributed to C1 and C2.  We are neither able to confirm that these sources have a tidal origin nor are primordial dark (or almost-dark) galaxies.

The relationship of C1 (and any possible interaction) with its \HI-rich neighbours is not obvious.  The ASKAP synthesis pre-pilot observations recover all of the \HI\ emission found from previous Parkes single-dish observations, and the ASKAP position is consistent with the Parkes single-dish centroid to within the positional uncertainties of the Parkes observations. 
Our ASKAP pre-pilot observations recover only 54~percent of C2's integrated \HI\ that was previously measured by the Parkes Basketweave survey.  Furthermore the position offset between the ASKAP and Parkes observations is in the direction of NGC~1385, within whose $r_{200}$ C2 is hypothesised to reside. In addition, the velocity of C2's \HI\ emission is the same as NGC~1385's recessional velocity. Hence if both C1 and C2 have a tidal origin and are candidate TDGs, then C1 is likely to be in a different stage of evolution than C2. 

In the hypothetical situation that one or both dark \HI\ sources are primordial dark galaxy candidates, we find from the TNG100 simulations that these sources likely formed in the early Universe more than 12~Gyr ago and had a relatively sedate evolutionary history whereby not much happened until the in-fall into a cluster environment 2~Gyr ago.  However, the observed differences between the \HI\ kinematics of C1 and C2, relative to those of the TNG-DGC suggests that more detailed modelling and more sensitive observations are required before any firmer conclusions can be made.

From a technical perspective, it is reassuring that these pre-pilot ASKAP synthesis observations have provided an improvement in terms of surface brightness sensitivity and angular resolution,  relative to past observations with the ATCA and during the early science phase of observations with ASKAP.  In this respect, the WALLABY survey will help further our understanding by providing a more complete census of the \HI\ Universe, as well as the low surface brightness `dark' Universe.

\section*{Acknowledgements}                                                     
                                                                                
We thank the anonymous referee for a constructive review of this paper.  Parts of this research was supported by the Australian Research Council Centre of Excellence for All-sky Astrophysics in 3 Dimensions (ASTRO 3D) through project number CE170100013. OIW acknowledges discussions with Elaine Sadler which helped improved this manuscript.   ARHS acknowledged receipt of the Jim Buckee Fellowship at ICRAR/UWA. SHO acknowledges a support from the National Research Foundation of Korea (NRF) grant funded by the Korea government (Ministry of Science and ICT: MSIT) (No. NRF-2020R1A2C1008706).  LVM and JR acknowledges financial support from the grants AYA2015-65973-C3-1-R and RTI2018-096228- B-C31 (MINECO/FEDER, UE),  as well as from the State Agency for Research of the Spanish MCIU through the "Center of Excellence Severo Ochoa" award to the Instituto de Astrofísica de Andalucía (SEV-2017-0709). JR acknowledges support from the State Research Agency (AEI-MCINN) of the Spanish Ministry of Science and Innovation under the grant "The structure and evolution of galaxies and their central regions" with reference PID2019-105602GB-I00/10.13039/501100011033.
FB acknowledges funding from the European Research Council (ERC) under the European Union’s Horizon 2020 research and innovation programme (grant agreement No.726384/Empire).  AB acknowledges support from the Centre National d'Etudes Spatiales (CNES), France.  PK is partially supported by the BMBF project 05A17PC2 for D-MeerKAT.  This work was supported by Funda\c{c}\~{a}o para a Ci\^{e}ncia e a Tecnologia (FCT) through the research grants UIDB/04434/2020 and UIDP/04434/2020. TCS acknowledges support from FCT through national funds in the form of a work contract with the reference DL 57/2016/CP1364/CT0009.

The Australian SKA Pathfinder is part of the Australia Telescope National Facility which is managed by CSIRO. Operation of ASKAP is funded by the Australian Government with support from the National Collaborative Research Infrastructure Strategy. ASKAP uses the resources of the Pawsey Supercomputing Centre. Establishment of ASKAP, the Murchison Radio-astronomy Observatory and the Pawsey Supercomputing Centre are initiatives of the Australian Government, with support from the Government of Western Australia and the Science and Industry Endowment Fund. We acknowledge the Wajarri Yamatji people as the traditional owners of the Observatory site.
 
This publication makes use of the NASA/IPAC Extragalactic Database (NED), which is operated by the Jet Propulsion Laboratory, California Institute of Technology, under contract with the National Aeronautics and Space Administration.

This project used public archival data from the Dark Energy Survey (DES). Funding for the DES Projects has been provided by the U.S. Department of Energy, the U.S. National Science Foundation, the Ministry of Science and Education of Spain, the Science and Technology FacilitiesCouncil of the United Kingdom, the Higher Education Funding Council for England, the National Center for Supercomputing Applications at the University of Illinois at Urbana-Champaign, the Kavli Institute of Cosmological Physics at the University of Chicago, the Center for Cosmology and Astro-Particle Physics at the Ohio State University, the Mitchell Institute for Fundamental Physics and Astronomy at Texas A\&M University, Financiadora de Estudos e Projetos, Funda{\c c}{\~a}o Carlos Chagas Filho de Amparo {\`a} Pesquisa do Estado do Rio de Janeiro, Conselho Nacional de Desenvolvimento Cient{\'i}fico e Tecnol{\'o}gico and the Minist{\'e}rio da Ci{\^e}ncia, Tecnologia e Inova{\c c}{\~a}o, the Deutsche Forschungsgemeinschaft, and the Collaborating Institutions in the Dark Energy Survey.  The Collaborating Institutions are Argonne National Laboratory, the University of California at Santa Cruz, the University of Cambridge, Centro de Investigaciones Energ{\'e}ticas, Medioambientales y Tecnol{\'o}gicas-Madrid, the University of Chicago, University College London, the DES-Brazil Consortium, the University of Edinburgh, the Eidgen{\"o}ssische Technische Hochschule (ETH) Z{\"u}rich, Fermi National Accelerator Laboratory, the University of Illinois at Urbana-Champaign, the Institut de Ci{\`e}ncies de l'Espai (IEEC/CSIC), the Institut de F{\'i}sica d'Altes Energies, Lawrence Berkeley National Laboratory, the Ludwig-Maximilians-Universit{\"a}t M{\"u}nchen and the associated Excellence Cluster Universe, the University of Michigan, the National Optical Astronomy Observatory, the University of Nottingham, the Ohio State University, the OzDES Membership Consortium, the University of Pennsylvania, the University of Portsmouth, SLAC National Accelerator Laboratory, Stanford University, the University of Sussex, and Texas A\&M University.  Based in part on observations at Cerro Tololo Inter-American Observatory, National Optical Astronomy Observatory, which is operated by the Association of Universities for Research in Astronomy (AURA) under a cooperative agreement with the National Science Foundation.  Database access and other data services are provided by the NOAO Data Lab.

\section*{Data Availability}
These pre-pilot observations are publicly-available from the CSIRO ASKAP Science Data Archive \citep[CASDA; ][]{chapman15,huynh20}.  Specifically, the visibilities and \HI\ data cubes for these observations can be found using this DOI:  {\url{https://dx.doi.org/10.25919/0yc5-f769}}

\bibliographystyle{mnras}
\bibliography{mn-jour,paperef}

\appendix
\section{Galaxy members of the Eridanus group}
\label{app-members}

\begin{table}
  \begin{scriptsize}
\begin{center}
\caption{Galaxy members of the Eridanus group as defined by \citet{brough06}.}
\begin{tabular}{lccc}
\hline
\hline
Galaxy & RA (J2000)& Dec (J2000) & $v$~(\kms) \\
\hline
ESO 482$-$G 017 & 03:37:43.33 &  $-$22:54:29.5&  1515 \\
LSBG F482$-$034 & 03:38:16.55 & $-$22:29:11.4 & 1359 \\
2MASX J03355395$-$2208228 & 03:35:53.95&  $-$22:08:23.0&  1374\\ 
2MASX J03354520$-$2146578 & 03:35:45.27 & $-$21:46:59.2 & 1638 \\
ESO 548$-$G 036 & 03:33:27.69 & $-$21:33:52.9 &  1520  \\
ESO 548$-$G 034 & 03:32:57.63 & $-$21:05:21.9 & 1707 \\
NGC 1353 &03:32:02.98  &$-$20:49:08.2 & 1587 \\
2MASX J03365674$-$2035231 & 03:36:56.75&  $-$20:35:23.0 & 1689 \\
NGC 1377 & 03:36:39.07 & $-$20:54:07.2&  1809 \\
ESO 548$-$G 069 & 03:40:36.17&  $-$21:31:32.4 &  1647 \\
NGC 1414 & 03:40:57.14 & $-$21:42:49.9 & 1752\\ 
APMUKS(BJ) B034114.27$-$212912 & 03:43:26.46 & $-$21:19:44.2&  1711 \\
NGC 1422 & 03:41:31.07 &  $-$21:40:53.5 & 1680 \\
NGC 1415 & 03:40:56.86 & $-$22:33:52.1 & 1659 \\
ESO 482$-$G 031 & 03:40:41.54 &  $-$22:39:04.1 & 1803 \\ 
ESO 548$-$G 029 & 03:30:47.17 & $-$21:03:29.6 & 1215 \\ 
IC 1953 & 03:33:41.87 & $-$21:28:43.1  &1867 \\
ESO 548$-$G 049 & 03:35:28.27 & $-$21:13:02.2&  1510\\ 
IC 1962 & 03:35:37.38 & $-$21:17:36.8 & 1806 \\
ESO 482$-$G 018 & 03:38:17.64 & $-$23:25:09.0 & 1687\\ 
NGC 1395 & 03:38:29.72 & $-$23:01:38.7 & 1717 \\
MCG $-$04$-$09$-$043 &03:39:21.57 & $-$21:24:54.6&  1588\\ 
NGC 1401 & 03:39:21.85 & $-$22:43:28.9 & 1495 \\
ESO 482$-$G 035 & 03:41:14.65 & $-$23:50:19.9 & 1890\\ 
NGC 1426 & 03:42:49.11 & $-$22:06:30.1&  1443 \\
ESO 549$-$G 006 & 03:43:38.25 & $-$21:14:13.7&  1609\\ 
NGC 1439 & 03:44:49.95 & $-$21:55:14.0 & 1670 \\
APMUKS(BJ) B033830.70$-$222643 & 03:40:41.35 & $-$22:17:10.5&  1737\\ 
ESO 482$-$G 027 & 03:39:41.21 & $-$23:50:39.8 & 1626 \\
ESO 548$-$G 070 & 03:40:40.99 & $-$22:17:13.4 & 1422 \\
ESO 482$-$G 036 & 03:42:18.80 & $-$22:45:09.2 & 1567 \\
\hline
\hline
\end{tabular}
\end{center}
\end{scriptsize}
\end{table}

\section{Cosmicflows-3 density field of the region between C1 and NGC~1403.}
\label{cosflows3}
\begin{figure}
\includegraphics[scale=.33]{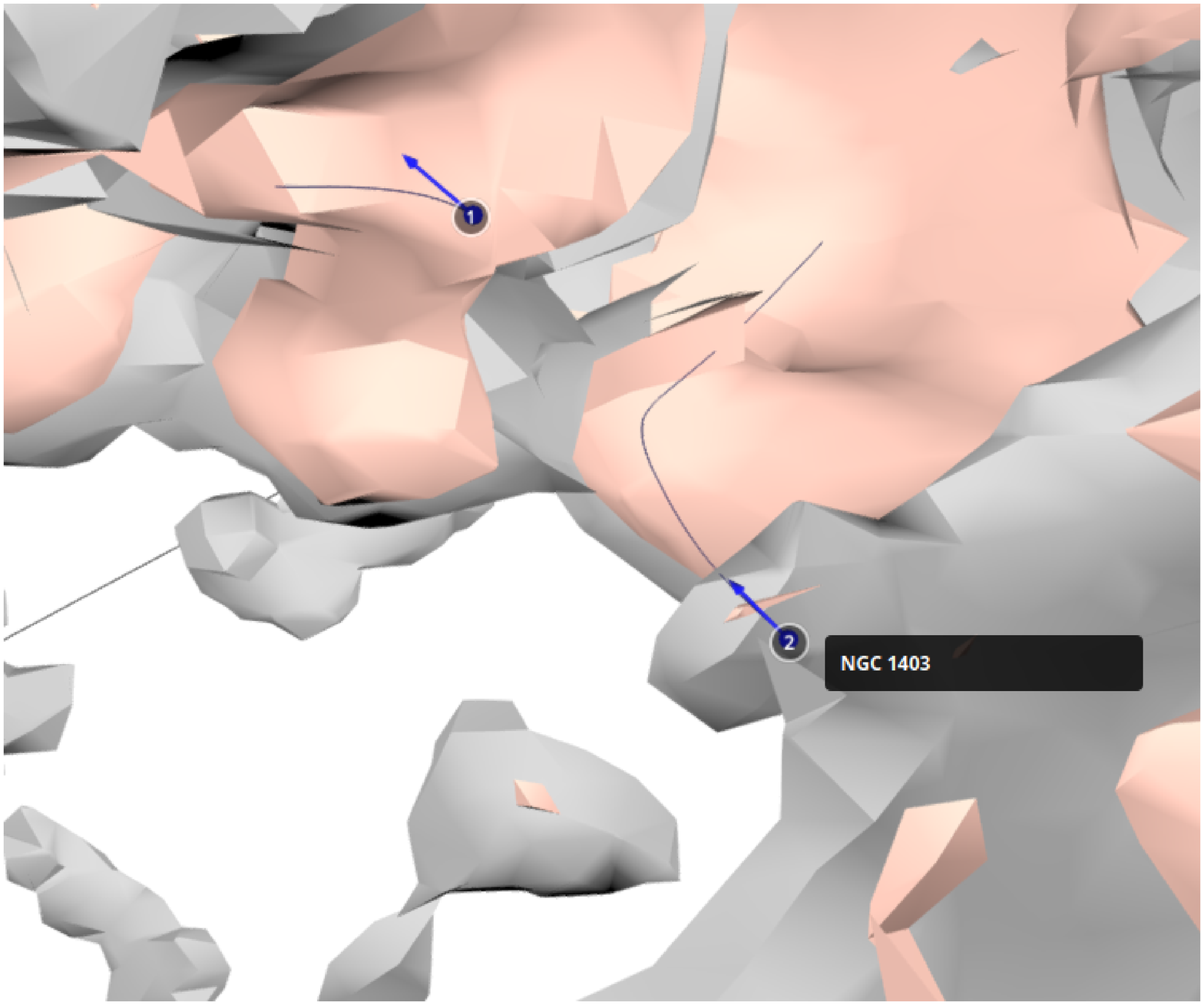}
\caption{The reconstructed density of the region between C1 (marked `1') and NGC~1403 (marked `2') from Cosmicflows-3 \citep{tully16}.  The blue vectors show the velocity trajectory of C1 towards the Great Attractor and NGC~1403 towards the Perseus-Pisces Supercluster. }
\label{cflow}
\end{figure}

\section{Comparing the 3D kinematic models for C2.}
\label{posvel}
\begin{figure}
\begin{tabular}{c}
\includegraphics[scale=.4,angle=270]{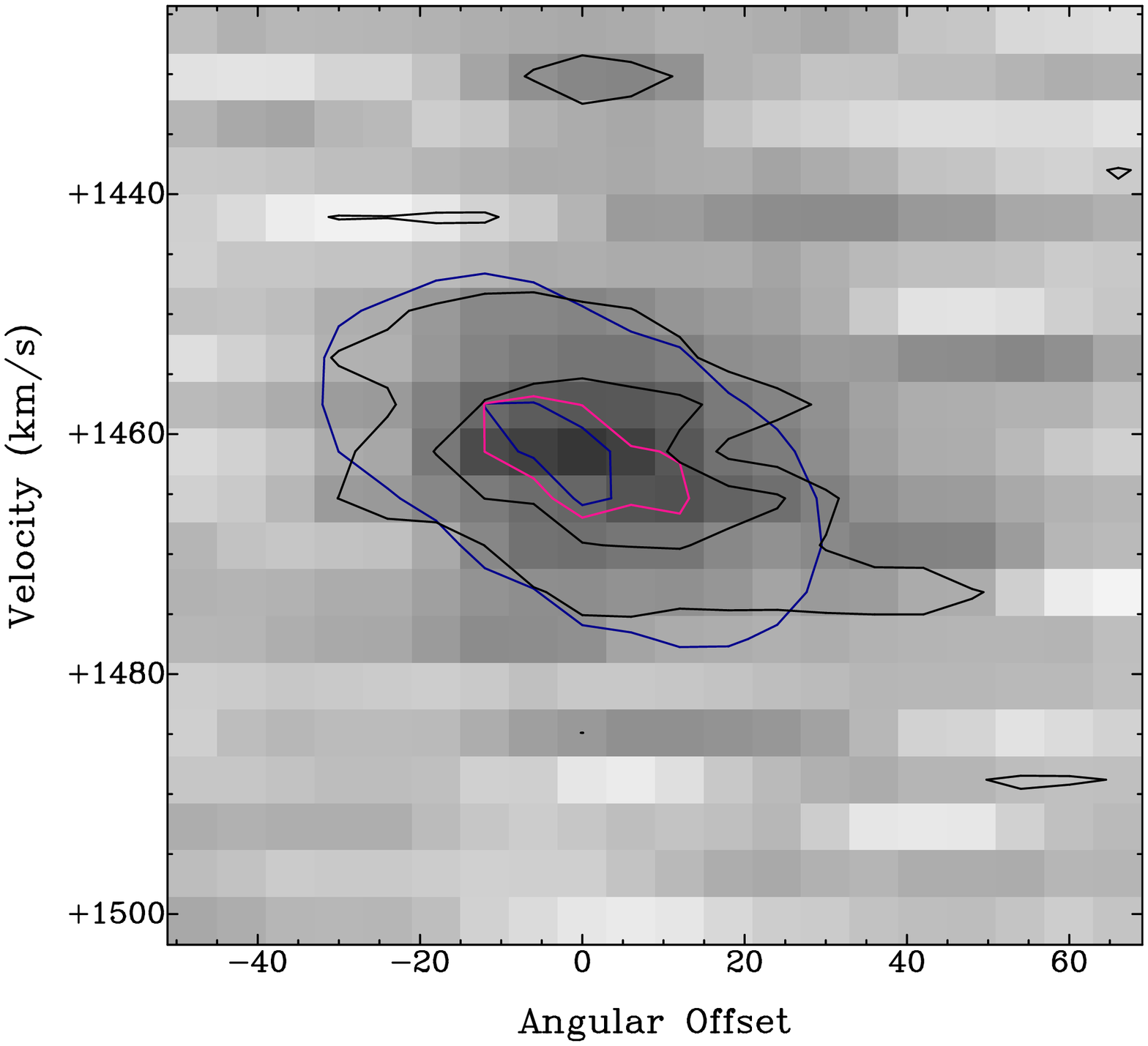}\\
\includegraphics[scale=.4,angle=270]{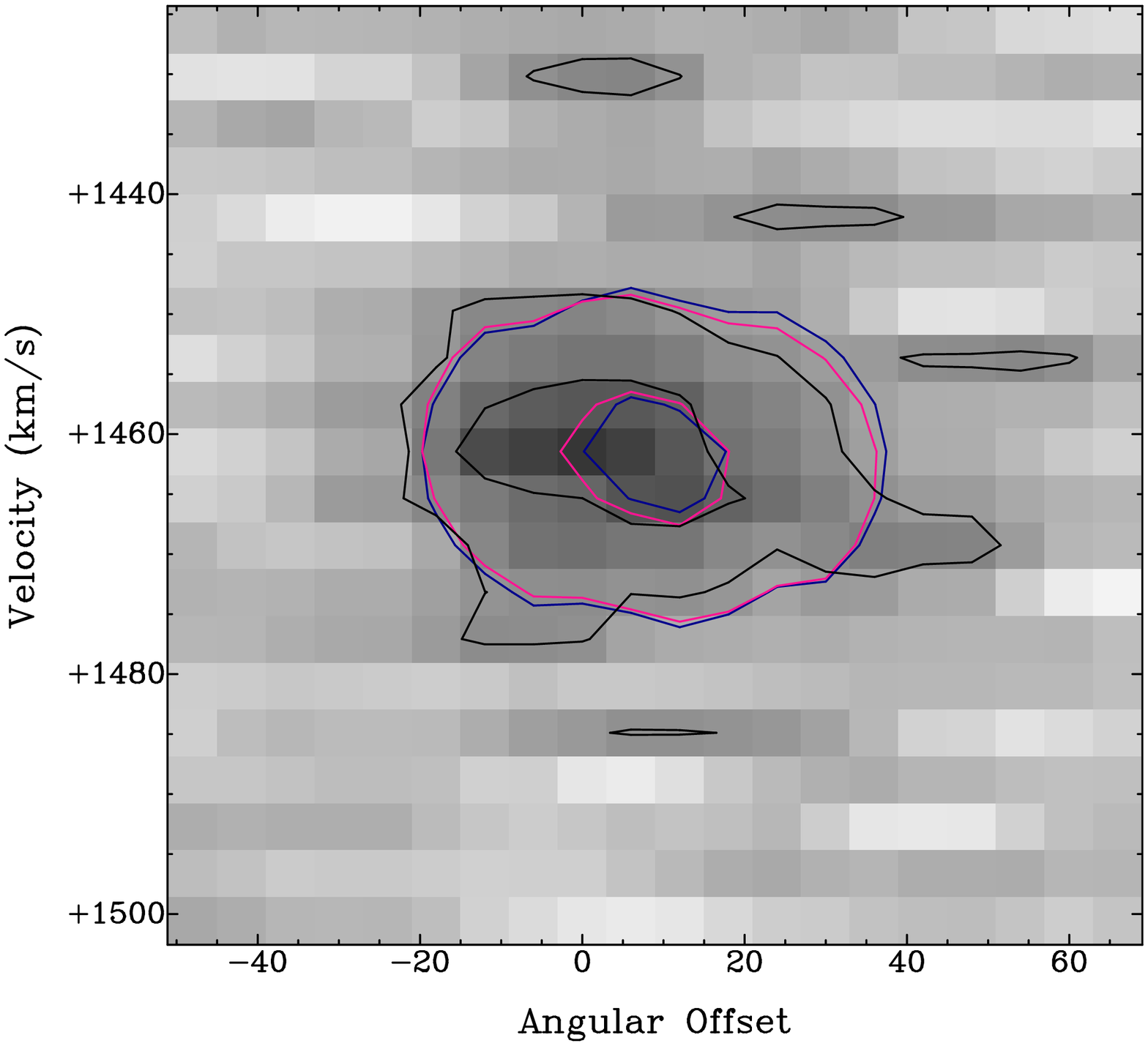}\\
\end{tabular}  
\caption{Position-velocity diagrams of C2 (black contour) along the major (top) and minor (bottom) axes of the modelled disk.  The `flat' model (where the inclination is allowed to vary) and the `edge-on' model (where the inclination is fixed at 90~$^{\circ}$) are represented by the blue and pink contours respectively.  We find that both 3D kinematic models fit the observations equally well despite the inclination differences --- a direct consequence of the low angular resolution of the observations.  }
\label{pvel}
\end{figure}

\label{lastpage}
\end{document}